%% file: main.tex
\newif\ifdoublecol
\doublecoltrue

\ifdoublecol
    \def \figwidth {0.95}
\else
    \def \figwidth {0.7}
\fi

\ifdoublecol
    \documentclass[dvipsnames]{IEEEtran}
\else
    \documentclass[12pt, draftclsnofoot, onecolumn,dvipsnames]{IEEEtran}
\fi
\usepackage[utf8]{inputenc}
\usepackage{amsmath,amsfonts, amssymb}
\usepackage{amsthm}
\usepackage{float, graphicx}
\usepackage{balance}
\usepackage{xcolor}
\usepackage{color,soul}
\usepackage{pgfplots}
\usepackage{standalone}
\usepackage{trdean}
\usetikzlibrary{arrows}
\pgfplotsset{compat=1.14}

\soulregister{\ref}{1}

\newcommand{\ftx}[1]{\mathcal{F}\left\{#1\right\}}
\newcommand{\invftx}[1]{\mathcal{F}^{-1}\left\{#1\right\}}

\begin{document}

\title{Rethinking Modulation and Detection \\ for High Doppler Channels}
\author{
Thomas Dean,~\IEEEmembership{Member,~IEEE,} Mainak Chowdhury,~\IEEEmembership{Member,~IEEE,}\\Nicole Grimwood,~\IEEEmembership{Member,~IEEE,} Andrea Goldsmith,~\IEEEmembership{Fellow,~IEEE}
\thanks{The authors are with the Department of Electrical Engineering, Stanford, CA
94305 USA (e-mail: trdean@stanford.edu, mainakch@stanford.edu, grimwood@alumni.stanford.edu, andrea@ee.stanford.edu).}%
\thanks{This work was presented in part in 2017 at IEEE PIMRC in Montr\'eal, Canada \cite{dean2017new}. T. Dean is supported by the Fannie and John Hertz Foundation. This work was supported in part by the NSF Center for Science of Information under Grant CCF-0939370.}}

\maketitle
\begin{abstract}
We present two modulation and detection techniques that are designed to allow for efficient equalization for channels that exhibit an arbitrary Doppler spread but no delay spread.  
These techniques are based on principles similar to techniques designed for time-invariant delay spread channels (e.g., Orthogonal Frequency Division Multiplexing or OFDM) and have the same computational complexity.  Through numerical simulations, we show that effective equalization is possible for channels that exhibit a high Doppler spread and even a modest delay spread, whereas equalized OFDM exhibits a strictly worse performance in these environments.  Our results indicate that, in rapidly time-varying channels, such as those found in high-mobility or mmWave deployments, new modulation coupled with appropriate channel estimation and equalization techniques may significantly outperform modulation and dectection schemes that are designed for static or slowly time varying multipath channels.

\vspace{1mm}
\noindent \emph{Index Terms} --- 5G Mobile Communication, Multipath Channels, Time-varying Channels, mmWave
\end{abstract}

\section{Introduction}
In many sub-6 GHz wireless systems such as LTE and WiFi, the delay spread of the point-to-point wireless channel is  typically much smaller than the coherence time \cite{goldsmith2005wireless, tse2005fundamentals}.  A long channel coherence time allows the wireless channel to be treated as a time-invariant channel with frequency-selective channel gains.  In this regime, OFDM, or general multicarrier modulation schemes, with rate and power adaptation, are well known to be information-theoretically optimal in terms of spectral efficiency \cite{gray2011entropy, durisi2010noncoherent}. In these channels, a long coherence time enables accurate channel estimation with negligible overhead.
Obtaining an accurate estimate of the channel becomes difficult as the coherence time decreases relative to the block length of the waveform, which can lead to significant performance degradation when using multicarrier modulation.

In general, the wireless channel is both time dispersive, introducing a delay spread, and frequency dispersive, introducing a Doppler spread \cite{matz2013time}.   
The capacity for a general time-varying channel with imperfect knowledge of and/or adaptation to the channel state is unknown. 
Channels with non-zero Doppler spread are no longer time invariant. Moreover, as the coherence time (roughly the inverse of the Doppler spread) shrinks relative to the block length, frequency-domain equalization methods, used with modulation and equalization techniques such as OFDM or Single-Carrier modulation with Frequency-Domain Equalization (SC-FDE), will no longer be effective as channel estimations will become inaccurate. Examples of wireless channels with significant Doppler spreads include mmWave systems and sub-6 GHz systems with high mobility \cite{rappaport2015wideband}.  Significant time-variations can also arise in non-terrestrial settings such as underwater systems \cite{dean2018underwater}, or in satellite-to-earth communication links \cite{zeng2016wireless}. Additionally, impairments such as phase noise in mmWave systems may manifest in ways similar to high Doppler spread in rapidly time-varying channels \cite{dhananhay2015thesis}.
The time duration over which the channel can be assumed to be time-invariant in such systems is much shorter than in typical sub-6 GHz systems;
these time-variations will affect the performance of algorithms that depend on accurate channel state information.

One approach to adapt modulation and detection techniques such as OFDM or SC-FDE to deal with the effects of time-variation is to limit symbol or block duration.
Indeed, is it typical for
OFDM deployments in time-varying channels to limit the symbol
time, or, equivalently, increase the subcarrier spacing, so that the
product of the overall symbol duration and the Doppler spread is
small.  In this regime, the time-frequency dispersive channel behaves approximately
as a static delay spread channel at the expense of an increased cyclic prefix overhead.  A detailed performance analysis of such
schemes is presented in \cite{wang2006performance}.  This class of
schemes will be further discussed in Section \ref{sec:simulation}.

A different approach to simultaneously deal with dispersion in the time and frequency domains
is via general time-frequency signaling \cite{matz2013time}.  
In \cite{liu2004orthogonal}, the authors describe a general framework for a time-frequency modulation scheme for time-frequency dispersive channels.  
The time-frequency representation in
\cite{liu2004orthogonal} uses the Short-time Fourier basis (SFT), and
proposes ways to deal with the loss of orthogonality between the basis
functions induced by the time-frequency dispersive channel.  
Another work, \cite{otfs2017hadani}, introduces Orthogonal
Time-Frequency-Space signaling (OTFS).  OTFS utilizes the
delay-Doppler representation of the signal and the channel. OTFS has been
demonstrated to have advantages over OFDM in specific
high-Doppler channels.  However, while \cite{otfs2017hadani} presents
an overall framework for designing waveforms, the performance of
OTFS in a time and frequency dispersive channel will depend on a large
number of tunable parameters.

In this work, we first consider a specific class of time-varying channels,
which are the time-frequency duals of time-invariant, frequency-selective channels.
Specifically these channels have zero delay spread and a finite Doppler spread.
We then describe two related modulation and detection schemes that are well suited to compensate for the impairments caused by these channels.  
We show that, provided a proper choice of waveform parameters and channel estimation algorithms, we may perfectly and efficiently compensate for the effects of this class of time-varying channels via equalization.  
Additionally, through numerical simulations, we show that these modulation and detection schemes perform well in channels that exhibit low to moderate delay spread and high Doppler spread.  We describe realistic terrestrial environments where such channels exist.

The modulation and detection schemes we present are the time-frequency duals of OFDM with Frequency-Domain Equalization\footnote{We refer to this modulation and detection technique as OFDM-FDE to emphasize that we are only considering OFDM with linear frequency-domain equalization as opposed to non-linear FDE or additional time-domain equalization.} (OFDM-FDE) and SC-FDE.  These proposed techniques, which we term Single-Carrier modulation with Time-Domain Equalization, or SC-TDE, and Frequency-Domain Multiplexing with a Frequency-Domain Cyclic Prefix, or FDM-FDCP, compensate for the effect of an arbitrary Doppler spread by performing linear equalization in the time domain\footnote{We note that FDM-FDCP could alternatively be called OFDM-TDE.  That is because, it is a form of OFDM with different equalization and channel estimation procedures.  However, we refer to this waveform as FDM-FDCP to highlight the necessity of the frequency-domain cyclic prefix in order to enable our equalization technique.  Additionally, the name FDM-FDCP remains consistent with existing literature on the waveform, e.g. \cite{dean2017new}.}.
This class of modulation and detection techniques is not degraded by arbitrary Doppler spreads because it directly estimates and equalizes the Doppler profile of the wireless channel rather than estimating and equalizing the delay-spread profile as is the case with OFDM-FDE.
Due to the fact that our techniques are the time-frequency duals of OFDM-FDE and SC-FDE, they inherit most of the
computational and implementation benefits found in these systems. 
\change{In fact, neglecting the inclusion of the frequency-domain cyclic prefix, the modulation of FDM-FDCP and SC-TDE are identical to
windowed versions of OFDM-FDE and SC-FDE. Our waveforms are differentiated by
their methods to
estimate and equalize the channel.}
The relationships between our techniques and OFDM-FDE and SC-FDE are shown in Figure \ref{fig:fig_duality}.

\ifdoublecol
\begin{figure*}
    \centering
    \begin{tikzpicture}[
    mod/.style=
        {rectangle, draw=none, rounded corners=1em,inner sep=8pt, minimum width=80pt},
    dualline/.style = 
        {
        -{Stealth[width=9pt,length=9pt]}, very thick, shorten >=12pt
        },
     dualliner/.style = 
        {
        {Stealth[width=9pt,length=9pt]}-, very thick, shorten >=17pt
        }
    ]
    \node at (5, 7.5) {Modulation Format};
    \node at (0, 6.5) {Multi Carrier};
    \node at (10,6.5) {Single Carrier};
    \draw (-2, 6) rectangle (12, -1);
    \node[rotate=90] at (-2.5, 0) {Frequency};
    \node[rotate=90] at (-2.5, 5) {Time};
    \node[rotate=90] at (-3.5, 2.5) {Equalization Domain};
    \node[mod, fill=blue!10] (fdmtdcp) at (0, 0) {OFDM-FDE};
    \node[mod, fill=blue!10] (tdmtdcp) at (10, 0) {SC-FDE};
    \node[mod, fill=red!10] (fdmfdcp) at (0, 5) {FDM-FDCP};
    \node[mod, fill=red!10] (tdmfdcp) at (10, 5) {SC-TDE};
    \draw[dualline] (tdmtdcp.north west) -- (fdmfdcp) node[near end, below=3pt] {$\mathcal{F}$};
    \draw[dualline] (fdmfdcp.south east) -- (tdmtdcp) node[near end, above=4pt] {$\mathcal{F}^{-1}$};;
    \draw[dualliner] (tdmfdcp.south west) -- (fdmtdcp) node[near end, above=2pt] {$\mathcal{F}^{-1}$};
    \draw[dualliner] (fdmtdcp.north east) -- (tdmfdcp) node[near end, below=3pt] {$\mathcal{F}$};
    \end{tikzpicture}    
    \caption{Relationships between existing modulation and detection techniques (blue) and our proposed techniques (red).  The scheme SC-TDE is the time-domain dual of OFDM-FDE whereas FDM-FDCP is the frequency-domain dual of SC-FDE.  For each technique, the cyclic prefix is placed in the dual domain, with equalization performed in that domain.  The benefits of channel estimation and equalization in these dual domains motivate our proposed schemes.}
    \label{fig:fig_duality}
\end{figure*}
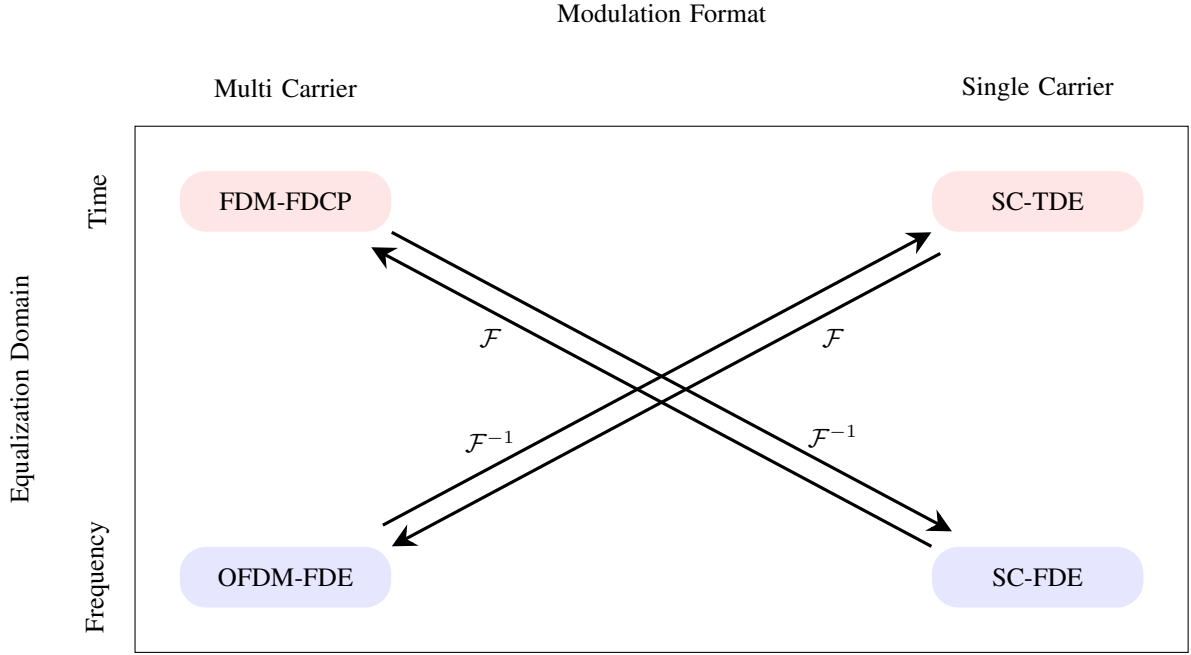
\fi

In addition to introducing this new class of modulation and detection schemes, we present several numerical simulations involving general time-varying, frequency selective channels.
We show several regimes where frequency-domain equalization schemes such as OFDM-FDE are strictly sub-optimal compared
to our techniques in terms of spectral efficiency and symbol error rates.  
Empirically, we find that our techniques are superior for channels with arbitrary Doppler spread provided the delay spread, $\tau_D$ is sufficiently small compared to the sampling rate, $T_s$. Specifically, we require $T_s \leq 2 / \tau_D$.
Conversely, in delay spread channels with no or little Doppler spread, our techniques perform worse than OFDM-FDE or SC-FDE. 
Our work suggests that rethinking common assumptions about joint waveform and equalizer design may result in significantly improved performance over current methods in many Doppler-spread channels of interest. Moreover, these proposed techniques have comparable computational complexities to existing methods.

The rest of the paper is organized as follows. In Section
\ref{sec:channel_model}, we describe notation and assumptions used throughout the paper.  In Section \ref{sec:fdcp}, we give a complete description of both the FDM-FDCP and SC-TDE modulation and detection techniques.  In Section
\ref{sec:simulation}, we present numerical results pertaining to various performance metrics and summarize the advantages and disadvantages of FDM-FDCP and SC-TDE compared to OFDM-FDE and SC-FDE.  Our
conclusions are presented in Section \ref{sec:conclusion}.

\section{System Model and Notation}
\label{sec:channel_model}
The complex baseband representation of the general time-varying impulse response function associated with a wireless channel is given by
 \begin{equation}
 \label{eq:impulse}
 h(t, \tau) = \sum_{i=0}^{K-1} \alpha_i e^{j 2 \pi f_i t} \delta(\tau - \tau_i),
 \end{equation}
 where $K$ is the number of multipath components and $\alpha_i, f_i, \tau_i$ are the complex gain, Doppler frequency and the delay associated with the $i^{th}$ multipath component, respectively.
This is an example of a channel introducing both time shifts and frequency shifts; an overview of this class of channels can be found in \cite{matz2013time}.  The channel is linear \textbf{time-invariant} \emph{only} when $f_i = 0$ for all $i,$ and the gains and delays associated with each individual component do not change in time.
Since we study only the effects of time-dispersion and frequency-dispersion, we assume henceforth that the delays, Doppler shifts and gains of the multipath components are constant over the transmit block duration.

Throughout this work, when we refer to a technique such as OFDM-FDE or FDM-FDCP, we are jointly considering modulation and detection, including the combination of channel estimation and equalization as part of the detection process.
Implicitly, this requires us to also make a set of assumptions about the nature of the channel (i.e. whether it is highly dispersive in time or frequency) in order to effectively measure and equalize the channel.
We precisely define how we perform channel estimation and equalization for our proposed techniques in Section \ref{sec:fdcp}.

 We assume that the time-varying impulse response function $h(t, \tau)$ is unknown at the transmitter and the receiver.  
The receiver performs  channel estimation based on pilot symbols transmitted using the modulation technique under consideration.  
For reasons further discussed in Section \ref{subsec:channel_estimation}, this means that different modulation and detection techniques will obtain different estimates for the channel with the same time-varying impulse response function  even in the absence of additive noise.
As a result, when used in differing classes of time-varying channel impulse response functions, for example with low or high Doppler or delay spreads, different techniques may have a substantially different SER performance, even without any additive noise 
(i.e., with infinite SNR).  We refer to channel estimates obtained in the limit of infinite SNR as \textbf{perfect} channel estimates.  

We note that for modulation and detection methods such as OFDM-FDE or SC-FDE, techniques exist that allow one to compensate for the effects of arbitrary Doppler spread.  For example, one may perform additional equalization in the frequency-domain, as described in \cite{guillaud2003channel}.  
However, such schemes require a complexity that is quadratic in the block length of the system, i.e. $O\left(N^2\right)$, and thus, in general are not well suited for real-time, high-throughput communications.  
In order to obtain a fair basis of comparison for all techniques presented in this paper, and also to provide constructions that are suitable for deployment in practical wireless systems, we only allow equalization that has an almost linear complexity, limiting the overall cost of  modulation and demodulation to the cost of the FFT operation, namely $O\left(N \log N\right)$.

We use the boldface notation $\mathbf{x}$ to denote a length $N$ discrete sequence, with $x[n]$ referring to the $n^{th}$ element of the sequence.
Unless otherwise specified, we use lowercase $\mathbf{x}$ to represent the time-domain sequence and uppercase $\mathbf{X}$ to denote its frequency domain representation.
The variable
$N$ refers to the block length of our waveform construction. 
Sequences of length $N$ are isomorphic to $N$-dimensional column vectors with complex entries in a natural way; hence we use $\mathbf{x}$ to refer to both a column vector and a finite length sequence of dimension $N$.  The operator $\ast$ represents linear convolution, the operator $\circledast$ represents circular convolution, and the operator $\odot$ represents the Hadamard product (element-wise scalar multiplication).  The symbols $\mathcal{F}$ and $\mathcal{F}^{-1}$ represent the DFT and inverse DFT operations respectively. The notation  $\frac{1}{\mathbf{x}}$ refers to a sequence $\mathbf{y}$ whose $n^{th}$ element is given by $y[n] = 1 / x[n]$, $\mathbf{x}^2 = \mathbf{x} \odot \mathbf{x}$, and $\mathbf{x}/\mathbf{y} = \mathbf{x} \odot \frac{1}{\mathbf{y}}$. Additive white Gaussian noise is denoted as $w(t)$ or $w[n]$.

\section{Modulation and detection for frequency dispersive channels}
\label{sec:fdcp} Techniques such as OFDM-FDE or SC-FDE, which rely on linear frequency-domain equalization, are designed for time-invariant delay spread channels.  Under time-invariance, the two-dimensional time-varying impulse response function reduces to a one-dimensional impulse response function, $h(\tau)$.  When the channel is not entirely time-invariant, $h(\tau)$ will not fully capture the effect of the two-dimensional channel.  Even if the channel estimate is obtained perfectly, i.e. in the absence of AWGN, the resulting equalization using this channel estimate may exhibit a residual error floor.  
We refer to this effect as \emph{model mismatch}. In Section \ref{sec:simulation}, we measure the effect of this model mismatch through the error vector magnitude (EVM) of the demodulated QAM symbols, measured in the absence of AWGN.

In order to study the potential gains that can be realized by changing our assumptions about the general behavior of the wireless channel,
we begin by considering the time-frequency dual of time-invariant delay spread channels.  
Specifically, we consider channels that have an arbitrary Doppler spread and zero delay spread.    
This assumption allows us to reduce the two-dimensional time-varying channel impulse response function to a different, one-dimensional function, namely
\begin{equation}
  \label{eq:dopp}
  h(t,\tau) = h(t) = \sum_{i=0}^{K-1} \alpha_i e^{j2 \pi f_i t}.
\end{equation}
For comparison, the corresponding one-dimensional channel response function for a time-invariant, non-zero delay spread, for which OFDM-like schemes are commonly used in many modern standards, is given by
\begin{equation}
  \label{eq:delay}
  h(t, \tau) = h(\tau) = \sum_{i=0}^{K-1} \alpha_i \delta(\tau - \tau_i).
\end{equation}
Notice that \eqref{eq:dopp} is a time-frequency dual of \eqref{eq:delay}.
At the receiver, the effect of the channel \eqref{eq:dopp} on the transmitted signal can be given by:
\begin{equation}
  \label{eq:dopp_cont}
  y(t) = \sum_{i=0}^{K-1} \alpha_i e^{j 2\pi t f_i} x(t).
\end{equation}
The continuous-time Fourier transform of this expression is given by
\begin{equation}
    Y(f) = \sum_{i=0}^{K-1} X(f - f_i),
\end{equation}
which is the linear convolution $H(f) \ast X(f)$, where $H(f)$ is the Fourier transform of the channel $h(t)$ given by {\eqref{eq:dopp}}.
Suppose that $X(f)$ is a strictly band-limited function with support $(-f_B, f_B)$.
Let $f_D \triangleq \text{max}_{i,j} |f_i - f_j|$ and consider a sampled version of $Y(f)$ over $(-f_B-f_D/2, f_B + f_D/2)$, with $N$ uniformly spaced samples with spacing of $\Delta f$.
Denote this length $N$ sequence as $Y[n]$.
Assuming that each $f_i$ lies on this sampling interval, then we have the following relation:
\begin{align}
  \label{eq:dopp_discr_fourier}
  \begin{split}
  Y[n] &= \sum_{i=0}^{K-1} \alpha_i X[n - f_i], \\
  \mathbf{Y}&= \mathbf{H} \ast \mathbf{X},
  \end{split}
\end{align}
where \[H[n] \triangleq H(f - n \Delta f) = \sum_{i=0}^{K-1} \alpha_i \delta[n - f_i]. \]  
Thus, the effect of the time-varying channel given by \eqref{eq:dopp} on the symbols $X[k]$ in the frequency domain is the same as the effect of the time-invariant channel given by \eqref{eq:delay} on time domain symbols.  
This implies that techniques used to correct for the delay spread in time invariant channels can be used to correct for the Doppler spread in time-varying channels.  
Exploiting this duality is the primary design idea behind the modulation and detection techniques that we present in this paper for high-Doppler spread channels.

We note that, in practice, the assumption used in \eqref{eq:dopp_discr_fourier}, that $f_i$ is discrete-valued, is not realistic.  
If our waveform is not properly constructed, this discrepancy can lead to large side-lobes in the frequency domain.
That is,
for a channel with $K$ multipath components, these sidelobes can cause a single symbol to interfere with more than $K$ symbols.   
We note that a similar effect, often termed ``tap leakage'', may also occur in OFDM-FDE- or SC-FDE-based systems when $\tau_i$ is not discrete-valued.  
This effect is discussed in \cite{van1995channel}.
In practice, this effect is mitigated by considering appropriate windowing or pulse shaping functions in the time and frequency domains.  
For SC-TDE and FDM-FDCP we may employ windowing functions which are approximately the duals of those commonly applied to OFDM-FDE and SC-FDE; this is described further in Section \ref{subsec:mcm_mod_demod}. 

SC-TDE is a single-carrier transmission scheme and carries QAM symbols in the time-domain; we denote the baseband sequence of QAM symbols as $r[n]$.  Conversely, FDM-FDCP is a multicarrier modulation scheme and contains QAM symbols in the frequency domain; the baseband QAM symbols of FDM-FDCP are denoted as $R[n]$.  
We note that when referring to SC-TDE signals, we use $R[n]$ to denote $\ftx{r[n]}$, and similarly, when referring to FDM-FDCP symbols, the sequence $r[n]$ denotes $\invftx{R[n]}$.

\subsection{Equalization and the frequency-domain cyclic prefix}
In the channel given by \eqref{eq:dopp_discr_fourier}, Doppler spread causes inter-carrier interference between frequency domain symbols that, if not corrected for, impairs performance and affects SER even without any additive noise.
As previously mentioned, one possible method to compensate for this interference is through equalization between frequency bins \cite{guillaud2003channel}.
More commonly, OFDM is adapted to time-varying channels by increasing the width of the subcarrier bin by reducing the overall symbol duration.  However, while this approach will reduce the ICI, the presence of Doppler will still lead to an error in channel estimation and may result in an increased symbol error rate.  This effect is explored numerically in Section \ref{sec:simulation}.

In the remainder of this section, we show how to compensate for the effect of Doppler spread through the use of a frequency-domain cyclic prefix.
For simplicity of exposition, we assume in this section that the receiver has perfect knowledge of the channel coefficients $h[n]$.
In Sections {\ref{subsec:channel_estimation}} and {\ref{subsec:mcm_mod_demod}}, we will return to describing how to perform channel estimation by placing appropriate pilot symbols in the frequency domain.
We first observe that in \eqref{eq:dopp_discr_fourier}, the effect of the channel is equivalent to linear convolution in the frequency domain.  This linear convolution may be converted to circular convolution by adding a cyclic prefix to the data symbols $\mathbf{X}$ in the frequency domain.
Similar to the time-domain cyclic prefix found in OFDM-FDE or SC-FDE,
the width of the cyclic prefix in the frequency domain depends on the Doppler spread of the channel. 
Specifically, if the Doppler spread in the discrete frequency domain is within the range $[-f_D / 2, f_D / 2],$ then the number of frequency domain cyclic prefix (FDCP) elements that need to be appended to $\mathbf{X}$ is $ f_D + 1$.

\begin{figure}
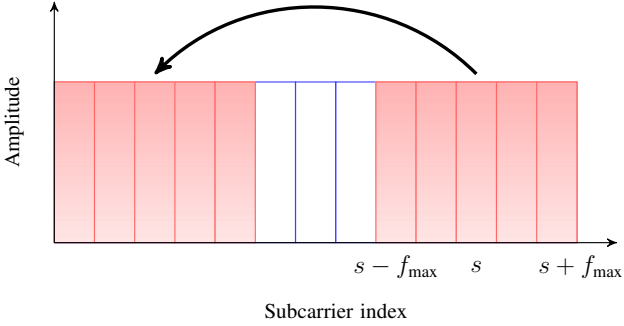

\centering
\includestandalone[width=\figwidth\columnwidth]{figures/fdcp_tikz}
\caption{Cyclic prefix operation in the frequency domain.  The last $f_D+1$ symbols are added to the beginning of the block \textbf{in the frequency domain}.}
\label{fig:freqcp}
\end{figure}

Denote $L \triangleq f_D + 1$, and assume that $R[n]$ is a sequence of length $N-L$ that contains data symbols with $L \ll N$. Let $\mathbf{\tilde{R}} = \{\tilde{R}[n]\}_{n=0}^{N - 1}$ be defined such that
\begin{equation}
  \label{eq:cp}
  \tilde{R}[n] =
  \begin{cases}
    R[n - L]  &\text{$n \geq L,$} \\
    R[N - L + n] &\text{otherwise}.
  \end{cases}
\end{equation}
This operation is illustrated in Figure \ref{fig:freqcp}. Denote $\mathbf{\tilde{r}} = \invftx{\mathbf{\tilde{R}}}$, of length $N$, as the time-domain sequence to be transmitted over the channel after pulse shaping.  
At the receiver, the stream of $N$ received symbols can be passed through a serial-to-parallel converter, and the resulting $L$ symbols can be removed from the frequency domain through the use of an FFT.  
The remaining $N - L$ symbols form the sequence $\mathbf{Z}$.  Ignoring additive noise, we have the following relation:
\begin{align}
\mathbf{Z} &= \mathbf{H} \circledast \mathbf{R}.
\end{align}
In the discrete frequency domain, the action of the channel can now be described as circular convolution, or equivalently as element-wise multiplication in the discrete time domain.  Equalization can thus be achieved by performing element-wise division by $\mathbf{h}$ in the time domain, i.e.
\begin{equation*}
    \hat{\mathbf{r}} = \invftx{\mathbf{Z}} / \invftx{\mathbf{H}} = \mathbf{z} / \mathbf{h}.
\end{equation*}

\change{
It is worth noting that, in a frequency dispersive channel that contains zero
delay spread, it is possible to equalize the effects of the channel through a
single-tap equalizer. Indeed, such an approach has been considered in works such
as \cite{bomfin2018theoretical}.  
Thus, one may be tempted to conclude that the inclusion of the FDCP is
unnecessary overhead. 
However, such a single-tap equalization scheme requires accurate knowledge of the channel and is highly sensitive to additional impairments such as non-zero delay spread. 
In contrast, the overhead of the FDCP is small, and its inclusion greatly
simplifies the process of equalization as well as channel estimation, as will be shown in Section {\ref{subsec:channel_estimation}}. 
A performance analysis comparison of FDM with FDCP versus with a single-tap equalizer in frequency dispersive channels is beyond the scope of
this work, as a comparison of this nature would need to incorporate estimation techniques and the associated error for the single-tap equalizer.
}

\subsection{Zero-Padding}
\label{subsec:zeropad}
In OFDM-FDE, the cyclic-prefix can be replaced by a guard interval in the time-domain equal to the length of the delay spread \cite{wang2000wireless}.  At the receiver, the resulting tail of each OFDM-FDE block falling into this guard interval is then added back to the beginning of the OFDM-FDE block, thus emulating the effect of the cyclic prefix.  This technique is known as zero-padded OFDM-FDE.

\begin{figure}
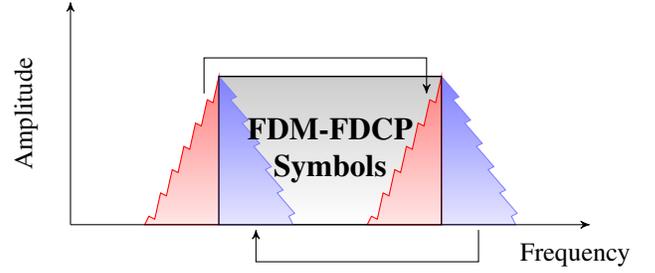

\begin{center}
\includestandalone[width=\figwidth\columnwidth]{figures/zero}
\end{center}
\par
\caption{Zero-padded FDM-FDCP:  Rather than including a cyclic-prefix in the frequency domain, we can simply add guard bands in the frequency domain, analogous to zero-padded OFDM-FDE.  The receiver performs equalization by adding the frequency-domain tails on both sides, represented here by shaded regions outside of the block labeled ``FDM-FDCP Symbols'', onto the original symbols and then performing element-wise division in the time domain. }
\label{fig:zero-pad}
\end{figure}

One may take a similar approach to emulate the frequency-domain cyclic prefix in FDM-FDCP.  In this case, at the transmitter, we must simply leave empty spectrum on both sides of the signal.  The receiver can then copy the tails in the frequency domain back into the FDM-FDCP block as shown in Figure \ref{fig:zero-pad}. This approach has only a minimal impact on the complexity at the receiver. 

A practical method of achieving zero-padding in a system transmitting FDM-FDCP is to simply place guard bands in the frequency domain.  In wireless systems, guard bands are already used to meet spectral mask requirements, and to help simplify filter design.
We note that unlike in the case of OFDM-FDE, the use of zero-padding over the full cyclic prefix results in a slightly reduced peak-to-average power ratio (PAPR).
This is discussed further in Section \ref{sub:papr}.

\subsection{Channel estimation}
\label{subsec:channel_estimation}
The measurement of the time-varying channel impulse response function, $h(t, \tau)$, is subject to the uncertainty principle \cite{matz2011fundamentals} arising from the fact that time and frequency are Fourier duals.  Given finite resources in time and bandwidth, there are fundamental limits to how accurately one can measure the delay and Doppler components of a given time-varying channel impulse response function.  
Any error associated with a measurement of the channel impulse response function will have an impact on the overall system performance of a modulation and detection scheme.

In OFDM systems, the assumption of time-invariance of the wireless channel helps with channel estimation.  
In the absence of Doppler spread, the channel impulse response is of the form \eqref{eq:delay} and is only a function of $\tau$.  
A non-zero delay spread gives rise to frequency selectivity;  
the frequency response, and hence the delay spread, can be  measured to any arbitrary precision by placing pilot symbols in the frequency domain. 

Equivalently, one may measure the entire frequency response of the channel by sending a single impulse in the time domain, which would correspond to sending an OFDM-FDE block with a constant data symbol in each frequency bin.
If the channel were truly time invariant, this estimate would then be valid for all future channel uses.
In practice, OFDM-FDE pilots are placed in the frequency domain and the channel gains are interpolated between pilots.
However, a variety of strategies exist to perform casual channel estimation, for example see \cite{van1995channel} or \cite{coleri2002channel}.  
For channels that are rapidly time-varying channels, these estimation strategies break down.  

\begin{figure}
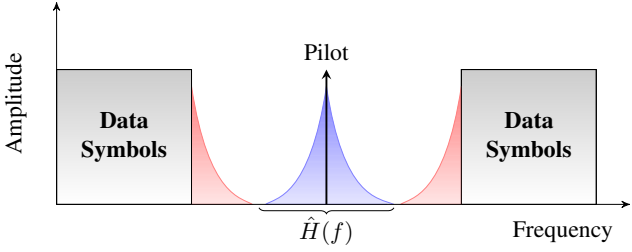

\begin{center}
\includestandalone[width=\figwidth\columnwidth]{figures/channel_estimation}
\end{center}
\par
\caption{A causal channel estimation strategy for FDM-FDCP.  A pilot tone may be placed in the frequency domain, surrounded by a guard interval at least as large as twice the Doppler spread.  The receiver may then recover an estimate of the Doppler profile, and hence the channel impulse response function, by simply measuring the resulting the received spectral components surrounding the pilot symbol. }
\label{fig:estimate}
\end{figure}

For FDM-FDCP and SC-TDE, channel estimation can be performed by assuming that a pilot sequence, $\mathbf{\tilde{r}} = \invftx{\mathbf{R}}$ is known at the receiver. 
An estimate  $\mathbf{\hat{h}} $ of the channel response $\mathbf{h} \triangleq \invftx{\mathbf{H}}$ can be calculated by
\begin{equation}
  \label{eq:channel_estimation}
   \mathbf{\hat{h}} = \invftx{\mathbf{Z}} / \invftx{\mathbf{R}}.
\end{equation}
Similar to the case for OFDM-FDE in the limit of zero Doppler spread, for FDM-FDCP with no delay spread, a full estimation may be performed by sending a single tone in the frequency domain, or transmitting a block with a constant single occupying every time slot.  
Equivalently, one may simply insert a pilot in the frequency domain and send no signal in an appropriate width surrounding the pilot, as shown in Figure \ref{fig:estimate}.
In the absence of delay spread, this channel estimate will remain valid over all blocks.
For SC-TDE, one may simply place pilots in the time domain and interpolate between these pilot symbols, in a manner analogous to performing estimation for OFDM-FDE.

Using FDM-FDCP, if the channel has a non-zero delay spread, the channel will become frequency selective.  
This implies that the channel estimate may differ depending on where the pilot is placed in the frequency domain.
The can be contrasted to the use of OFDM-FDE in a high-Doppler environment where the channel will vary over the length of the block, making the channel estimate inaccurate.
In OFDM-FDE, the time-duration of the block length is limited by the coherence time of the channel, which is inversely proportional to the Doppler spread. 
In contrast, FDM-FDCP is limited in \emph{bandwidth} by the coherence bandwidth of the channel, which is inversely proportional to the delay spread of the channel.
The effectiveness of our modulation and detection techniques at equalizing channels with non-zero delay spread will be investigated further through numerical simulations in Section \ref{sec:simulation}.

\subsection{Modulation and detection}
\label{subsec:mcm_mod_demod}

\ifdoublecol
\begin{figure*}
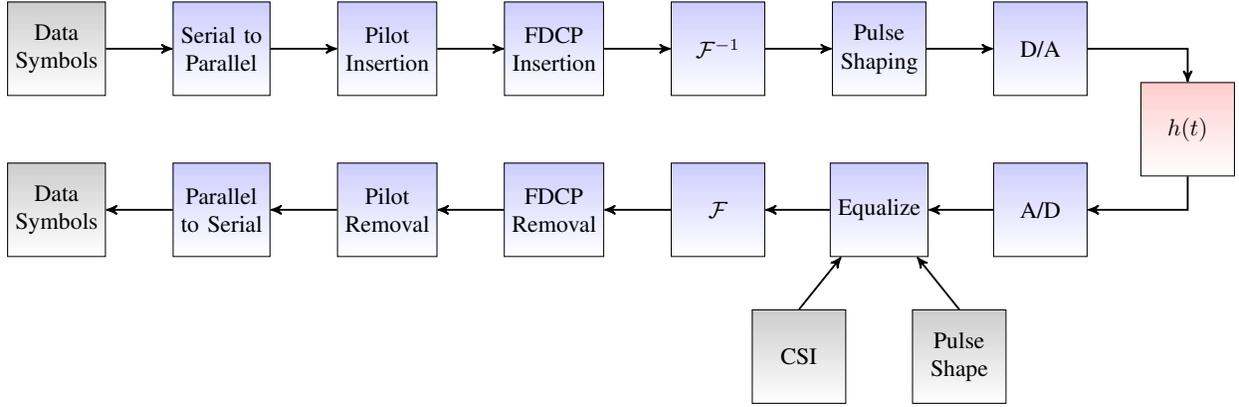
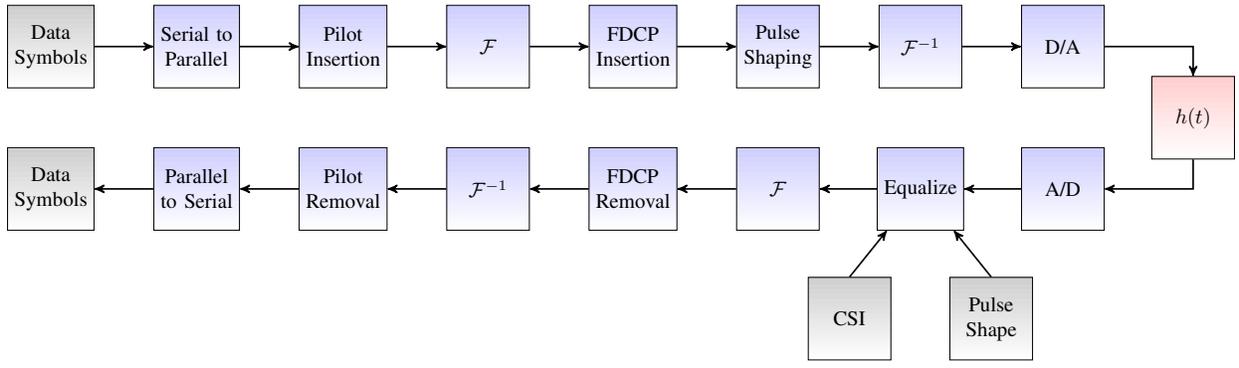

\centering
\subfloat[FDM-FDCP]{
    \includestandalone[width=0.9\textwidth]{figures/system_model}
}
\vspace{1cm}

\subfloat[SC-TDE]{
    \includestandalone[width=0.9\textwidth]{figures/system_sc_tde}
}
\caption{(a) A system diagram of FDM-FDCP.  Pilot symbols and a frequency-domain cyclic prefix are added to the data symbols, which are then modulated through an inverse FFT.  Pulse shaping is performed in the time domain analogous to OFDM-FDE.
SC-TDE is shown in (b), which differs from FDM-FDCP through an additional Fourier transform.  Transmit pulse shapes are windowed in the frequency domain in analogy to SC-FDE. For both systems, the FFT and IFFT are the most expensive operations from the perspective of computational complexity and require $O(N \log N)$ operations.}
\label{fig:system_model}
\end{figure*}
\fi
Having described the operations of channel estimation and equalization, we now fully describe how to modulate and detect FDM-FDCP and SC-TDE.  Overall block diagrams for both systems are presented in Figure \ref{fig:system_model}.
At the transmitter, in both systems, QAM symbols are converted from serial to parallel blocks of length $N-L-P$, where $P$ is the number of pilot symbols, including any guard intervals associated with the pilot if applicable.  Pilot tones are then inserted along side the data symbols as described in Section \ref{subsec:channel_estimation}.  
For FDM-FDCP, a single tone may be placed in the center of the signal bandwidth surrounded by appropriate guard bands.  
For SC-TDE, pilots are interleaved with the time-domain data analagous to OFDM-FDE.
The frequency domain cyclic prefix is then appended to the waveform, and
the resulting signal is then passed through an appropriate pulse shaping or windowing function before being transmitted over the air.

It is worth noting that the modulation and demodulation of FDM-FDCP are less expensive computationally than SC-TDE, as the modulation and demodulation of SC-TDE requires an extra FFT and IFFT operation respectively.
In addition, FDM-FDCP is similar to OFDM-FDE in
implementation\footnote{\change{In fact, disregarding the frequency-domain
cyclic prefix, the modulation of FDM-FDCP is
identical to that of windowed OFDM.
However, our proposed modulation differs due to the inclusion of the FDCP in the FDM modulation, as well as a
different process for channel estimation and equalization.
}}.
For these reasons, one may find FDM-FDCP a more natural replacement for OFDM-FDE. 
However, as further discussed in Section {\ref{sec:simulation}}, if CSI is known at the transmitter, one may apply adaptive rate and power allocation to SC-TDE, and thus, a higher achievable rate may be possible using SC-TDE rather than FDM-FDCP.

For this work, we consider only a single pulse shape and attempt to use equivalent pulse-shaping methods to compare each modulation and detection technique.
A more complete description on the effect of pulse shaping on our techniques is beyond the scope of this work.
However, we note that due to the similarity of our techniques to existing constructions, existing work on pulse shaping for OFDM-FDE or SC-FDE may be applied, see for example \cite{bolcskei1999design} or \cite{schafhuber2002pulse}.
For each modulation technique, pulse shaping is applied at the transmitter in the domain that is dual to the data symbols.  Specifically, we rely on the standard root raised cosine response, $G(s)$, given by
\begin{equation}
    G(s) =
    \begin{cases}
    \sqrt{T} & 0 \leq |s| \leq \frac{1 - \beta}{2 T} \\
    \sqrt{\frac{T}{2} \left(1 + \cos \left[ \frac{\pi T}{\beta} \left(|s| - \frac{1-\beta}{2T} \right) \right] \right) } & \frac{1-\beta}{2T} \leq |s| \leq \frac{1+\beta}{2T} \\
    0 & |s| > \frac{1+\beta}{2T}.
    \end{cases}
\end{equation}
For FDM-FDCP, we multiply the \emph{time-domain} signal by $G(t)$ with $\beta = 0.1$ and $T = T_s$.  This yields orthogonal sub-carriers that consist of Nyquist-like pulses with a rapidly decaying tail.  For SC-TDE, we multiply the \emph{frequency-domain} signal by $G(f)$ with the same parameters so that the resulting time-domain signal is a Nyquist pulse with 10\% excess bandwidth.
In the simulations presented in Section {\ref{sec:simulation}}, these operations are accomplished by resampling the baseband signal at a rate of 5/4 and then filtering with an appropriately discretized version of $G(t)$ or $G(f)$.  
These operations may be performed efficiently through the use of a polyphase filter, for example.

Since channel estimation is performed by transmitting pilots through the appropriate pulse-shaping function, we may define the effective channel response as the composition of the pulse-shaping operator and the channel operator.  That is, assuming our channels have zero delay spread\footnote{For the general channel with non-zero delay and Doppler spread, the convolution and multiplication operators are replaced with more general compositions for linear operators.}, $h_{\text{eff}}[n] = h[n] \odot G[n]$ for FDM-FDCP and $h_{\text{eff}}[n] = h[n] * g[n]$, for SC-TDE, where $g[n] = \invftx{G[n]}$.
Then we can follow exactly the same steps as in \eqref{eq:channel_estimation} for channel estimation and data demodulation.  If $y[n]$ is the signal received after discretization at the receiver, and assuming the receiver has recovered an estimate of the channel, $\hat{h}_{\text{eff}}[n]$ we may estimate $\tilde{r}[n]$ as
\begin{equation}
    \label{eq:channel_estimation_pulse}
    \tilde{r}[n] = y[n] / \hat{h}_{\text{eff}}[n].
\end{equation}
Finally, an estimate of the data symbols may be recovered by removing the frequency-domain cyclic prefix and pilot symbols in their appropriate domains.  We note that computing the Fourier transform and inverse-Fourier transform of a block of symbols is the most computationally expensive operation and requires $O(N \log N)$ operations.

\subsection{Peak-to-Average Power Ratio Considerations}
\label{sub:papr}
Multicarrier modulation schemes such as OFDM often have a large peak-to-average power ratios (PAPR). 
This may lead to difficulties in the implementation of such schemes due to non-linearities present in transmitter power amplifiers.
We note that due to the similarity of our waveforms to existing constructions, it should be possible to apply many PAPR reduction techniques developed for OFDM to our waveforms, for example see \cite{tellado2006multicarrier}.

In general, the peak-to-average power ratios of the transmitted FDM-FDCP and SC-TDE waveforms are very similar to those of OFDM-FDE and SC-FDE respectively.  
We expect that the overhead of the frequency-domain cyclic prefix will cause a slight increase in PAPR compared to schemes that use only a time-domain cyclic prefix. 
That is, we expect FDM-FDCP will have a slightly higher PAPR than OFDM-FDE.
The SC-TDE waveform will have a lower PAPR due to the fact that the data symbols are transmitted in the time domain.  
However, the application of the frequency-domain cyclic prefix will increase the PAPR beyond that of the SC-FDE waveform.
As the analytic study of PAPR can often be difficult, a more complete discussion of PAPR is beyond the scope of this work.

\section{Numerical Results}
\label{sec:simulation}

We now present a series of simulations that show the potential of both FDM-FDCP and SC-TDE to mitigate the effects of Doppler spread. The channel model used in these simulations is described in detail in Section \ref{sub:channel_models}.
In order to highlight the effect of model mismatch, we assume that a perfect, non-casual channel estimate is available at the receiver.
As described in Section \ref{sec:fdcp}, this estimate is not the two-dimensional scattering function, but rather $h_{\text{eff}}[n]$ which captures the effect of using a pulse shaping filter.
In our simulations, $h_{\text{eff}}[n]$ is obtained by transmitting an appropriate pilot sequence through the channel given by $h(t,\tau)$ at time $t=0$ without the presence of AWGN using the process described in Section {\ref{subsec:channel_estimation}}.
A data sequence is then transmitted through the channel $h(t,\tau)$ at time $t=0$ and equalized according to $h_{\text{eff}}[n]$.
In practice, the receiver will not have access to such a non-causal estimate of the channel, and the pilot sequence will need to be placed within a data block, as described in Section {\ref{subsec:channel_estimation}}.  
This allows us to decouple channel estimation errors (due to noise, suboptimal channel estimation schemes, etc.) from errors due to the model mismatch introduced by the waveform.  
In practice, both data transmission and  channel estimation will be interleaved; optimizing both of them jointly is an open problem for general channels and is outside the scope of this paper.

In order to quantify the effects of model mismatch, we rely on the error-vector magnitude (EVM) metric, which is defined as the ratio of the amplitude of the error vector to the root mean squared amplitude of the received symbol, or
\begin{equation}
    \text{EVM} = \frac{\sqrt{P_{\text{error}}}}{\sqrt{P_{\text{mean}}}}.
\end{equation}
It is also convenient to express decoding error in terms of an irreducible-error floor.  If we assume that the error-vector is approximately Gaussian, then using MQAM modulation, we can approximate the irreducible-error floor in terms of SER using the standard expression for the SER performance of MQAM in AWGN (e.g. see \cite[Chapter 6]{goldsmith2005wireless}), namely
\begin{equation}
    P_{s,\text{floor}} = \frac{2(\sqrt{M}-1)}{\sqrt{M}} \, Q \left( \sqrt{ \frac{3}{(M-1)(\text{EVM})^2} } \right).
\end{equation}

We note that our simulations show little difference in performance between single and multicarrier modulation formats (i.e. SC-FDE vs. OFDM-FDE or SC-TDE vs. FDM-FDCP).  
This is because we do not assume CSI knowledge at the transmitter and therefore there is no rate or power adaptation across subcarriers or time-domain symbols. 
For SC-FDE, if CSI is known at the transmitter, the variable-rate and variable-power QAM modulation scheme described by \cite{goldsmith1997variable} may be employed.
Further performance differences between single and multicarrier schemes may be
observed in channels with deeper fading characteristics, or when using
error-correcting codes.\change{  
In particular, schemes using SC-FDE combined with error correcting codes and turbo equalization have shown to substantial gains over multi-carrier modulation formats
\cite{ye2007frequency}}.

In all simulations, we use the zero-padded version of the waveform, discussed in Section {\ref{subsec:zeropad}}.
For OFDM-FDE and SC-FDE, a cyclic prefix (or guard interval) of length $\text{min}( \lceil 5 \, \tau_{\text{RMS}} \, F_s \rceil, 8 )$ is appended to the end of each transmit block.  For FDM-FDCP and SC-TDE, a cyclic prefix (or guard band) of length $\text{min}( \lceil 2 N T_s f_D \rceil, 8 )$ is appended on both sides of the transmit spectrum.
For FDM-FDCP and SC-TDE, the cyclic prefix is padded beyond its required length to ensure we account for the formation of sidelobes caused by Doppler components that do not fall on a frequency bin.
For OFDM-FDE and SC-FDE, the length of the time-domain cyclic prefix is based on the RMS delay spread and chosen to be sufficiently long so that received power falling outside of the cyclic-prefix interval is at least 20 dB below the averaged received power.
For all simulations, we have found that further increasing the cyclic prefix length results in a negligible change in performance.

\subsection{Channel Models}
\label{sub:channel_models}
The channel models used in these simulations are based on the 3GPP channel models for 0.5 to 100 GHz found in TR 38.901  \cite{3gppChannel}.  
To emphasize the effects of delay and Doppler spread on the modulation format, we choose a non-line-of-sight link between a single transmitter and a single receiver.  
Specifically, we consider that each channel is drawn according to the two-dimensional channel scattering function
\begin{equation}
    h(t,\tau) = \sum_{i=1}^n \alpha_i e^{2 \pi f_i t} \delta( \tau - \tau_i ).
\end{equation}
For a specified delay spread, the power and delay of each received component is fixed according to the Tapped-Delay Line Model-A (TDL-A); the normalized values of $|\alpha_i|$ and $\tau_i$ are given in Table \ref{tab:tdla}.
\begin{table}
    \centering
    \caption{TDL-A model parameters}
    \begin{tabular}{@{}l l l|l l l@{}}
    \toprule
        Tap & Normalized &  Power & Tap & Normalized & Power \\
        Number & Delay & (dB) & Number & Delay &  (dB) \\ \hline
        1 & 0      & -13.4  & 12 &  2.2242 & -16.7\\
        2 & 0.3819 & 0      & 13 & 2.1718 & -12.4 \\
        3 & 0.4025 & -2.2   & 14 & 2.4942 & -15.2 \\
        4 & 0.5868 & -4     & 15 & 2.5119 & -10.8\\
        5 & 0.4610 & -6     & 16 & 3.0582 & -11.3\\
        6 & 0.5375 & -8.2   & 17 & 4.0810 & -12.7\\
        7 & 0.6708 & -9.9   & 18 & 4.4579 & -16.2\\
        8 & 0.5750 & -10.5  & 19 & 4.7966 & -18.3 \\
        9 & 0.7618 & -7.5   & 20 & 5.0066 & -16.6\\
        10 & 1.5375 & -15.9 & 21 & 5.3043 & -19.9\\
        11 & 1.8978 & -6.6  & 22 & 9.6586 & -29.7 \\
    \bottomrule
    \end{tabular}
    \label{tab:tdla}
\end{table}
For each simulation, each received component is assigned a phase uniformly at random. Further,
each $f_i$ was drawn according to a Jakes' spectrum \cite{jakes1974microwave}; that is, each $f_i$ is a random variable that is drawn independently according to the PDF
\begin{equation}
p(f) =
\begin{cases}
\frac{1}{\pi f_D \sqrt{1 - (f/f_D)^2}}, &  |f| < f_D \\
0, & \text{otherwise}.
\end{cases}
\end{equation}
All channel parameters are assumed to be constant over each transmission block length.

The total channel model can now be fully characterized by specifying only the desired delay and Doppler spread.  
We note that a specified Doppler spread is the maximum frequency difference between any two received components, that is, $f_D = \max_{i,j} |f_i - f_j|$. 
In contrast, as shown in Table \ref{tab:tdla}, the reported delay-spread is an RMS delay spread, and the received components will have a maximum delay that is nearly ten times the delay specified by the delay spread; however, over 90\% of the signal power will be captured within the time interval indicated by the delay spread.

\begin{table}
    \centering
    \caption{Delay and Doppler profiles used in this work and corresponding deployment scenarios.}
    \begin{tabular}{@{}llll@{}}
          & \multicolumn{3}{c}{$f_c$} \\
          \cmidrule{2-4}
          $f_D$   & 900 MHz & 6 GHz & 60 GHz \\ \toprule
         50 Hz  & $\pm 30$ kmh & $\pm 5$ kmh & $\pm 0.5$ kmh \\
         500 Hz & -- &  $\pm 50$ kmh &   $\pm 5$ kmh \\ 
         $5\,000$ Hz & -- & -- & $\pm 50$ kmh \\
     \bottomrule
    \end{tabular}
    \ifdoublecol
        \vspace{5 mm}
        
    \fi
    \begin{tabular}{@{}lll@{}}
         \toprule
         Normalized & Description & Scenario \\
         Delay & & \\ \hline
         10 ns  & Very short & \multirow{2}{*}{Indoor Office} \\
         20 ns  & Normal & \\ \hdashline
         50 ns & Short & \multirow{2}{*}{Urban Canyon} \\
         100 ns & Normal & \\ \hdashline
         300 ns & Normal & \multirow{2}{*}{Urban Macro} \\
         $1\,000$ ns & Long & \\
     \bottomrule
    \end{tabular}
    \label{tab:Doppler}
\end{table}

The range of delay and Doppler spreads used in our simulations correspond to realistic wireless environments.  
In particular, for delay spread we consider a minimum of 10 ns, which in \cite{3gppChannel} is described as a ``very short'' delay spread, and is typical of an indoor or office environment.  We also consider delay spreads as high as 100 ns  which \cite{3gppChannel} describes as a ``normal'' delay spread and would be typical in an urban canyon deployment, for example.  In Table \ref{tab:Doppler}, we list typical deployment scenarios corresponding to differing delay spread for 6 GHz deployments.
We note that higher-frequency deployments such as 20 or 60 GHz would experience delay spreads slightly lower than indicated by Table \ref{tab:Doppler}.

\begin{figure}
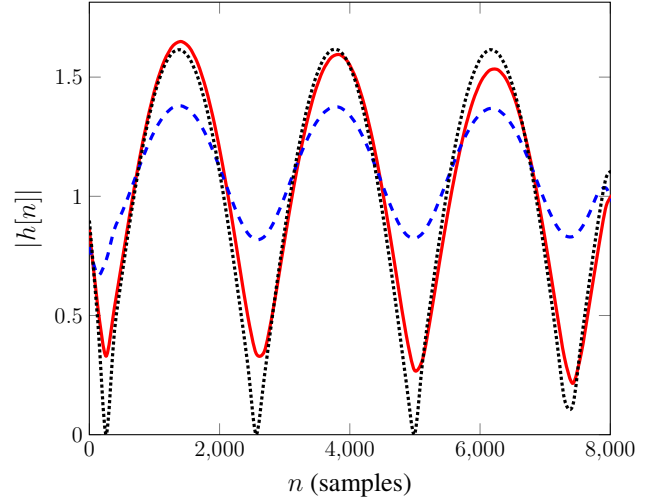

    \centering
    \includestandalone[width=\figwidth\columnwidth]{figures/fading}
    \caption{Representative time-domain channel responses, measured through the FDM-FDCP estimation process described in Section \ref{subsec:channel_estimation}.  The channel considered has a 500 Hz Doppler spread and a 100 ns delay spread where delays components are drawn from the TDL-A model and Doppler components are drawn from the Jakes' model.  The sampling rate is 4.5 MHz.}
    \label{fig:fading}
\end{figure}

The Doppler spread values considered range from 50 Hz to $5\,000$ Hz. 
We note that Doppler spreads as high as 500 Hz would be typical for a 60 GHz deployment in an environment with low mobility (walking speeds), and would arise in 6 GHz deployments with vehicles moving at surface-road speeds.
Table \ref{tab:Doppler} lists scatterer speeds relative to either transmitter or receiver that result in 50 or 500 Hz Doppler shifts for various center frequencies.
Several representative channel responses obtained using this model are presented in Figure \ref{fig:fading}.
These responses are measured through the FDM-FDCP estimation routine described in Section \ref{subsec:channel_estimation}.

\subsection{Decoding Performance under AWGN}
To demonstrate the ability of our modulation and detection techniques to compensate for arbitrary Doppler spread, we simulate the decoding performance under AWGN for channels with various Doppler spreads.  The first set of simulations are given in Fig. \ref{fig:awgn} where, for comparison, we also provide a simulation of OFDM-FDE. 
For both techniques, we fix the blocklength to be $N=2\,048$ with a sampling rate of 1.92 MHz.  
We simulate transmitting 16 QAM over all channels.
In order to emphasize channel impairments caused by Doppler, we choose a small delay spread of 50 ns, typical of an indoor setting or a mmWave urban-canyon deployment.
For both techinques, we assume that CSI is known perfectly at the receiver.  The results in this figure are averaged over $100\, 000$ channel realizations.  
In each channel realization, the delay taps and amplitudes remain fixed and are drawn in accordance with the TDL-A model, but the Doppler shift associated with each scatterer is re-drawn according to the Jakes' model. 
In addition, each scatterer is assigned a uniformly random phase rotation in each channel realization.

\ifdoublecol
\begin{figure*}
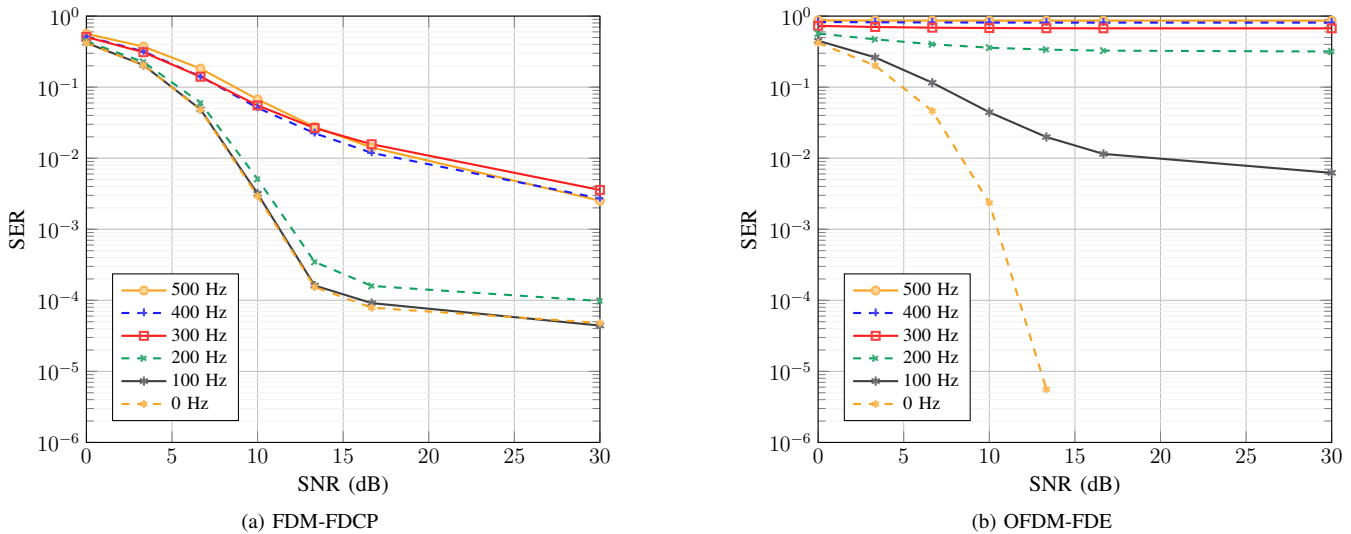

    \centering
    \subfloat[FDM-FDCP\label{sub:FDM-FDCP}]{
        \includestandalone[width=0.45\textwidth]{figures/ser_fdm_fdcp}
    }
    \hfill
    \subfloat[OFDM-FDE\label{sub:OFDM-FDE}]{
    \includestandalone[width=0.45\textwidth]{figures/ser_ofdm}
    }
    \caption{(a) The decoding performance of 16 QAM using FDM-FDCP with a 1.92 MHz sampling rate and a block length of $N=2\,048$ under AWGN.  The channel model is based on the 3GPP TDL-A model with a normalized delay spread of 50 ns. (b) OFDM-FDE in the same channel using the same set of parameters.  For both techniques, we assume perfect CSI is available at the receiver. The results in these figures are averaged over $100\, 000$ channel realizations.}
    \label{fig:awgn}
\end{figure*}
\fi

For small Doppler spreads, i.e. 200 Hz and below, FDM-FDCP is able to almost entirely equalize the channel.  As the Doppler spread increases, the small amount of delay spread somewhat impairs our ability to fully equalize the channel.  
This effect is considered more carefully in Section \ref{sub:channel_sweep}.
In contrast, OFDM-FDE is unable to equalize the channel except when the Doppler spread is below 100 Hz; however, even in this case OFDM-FDE has substantially worse SER performance than FDM-FDCP.

\ifdoublecol
\begin{figure*}
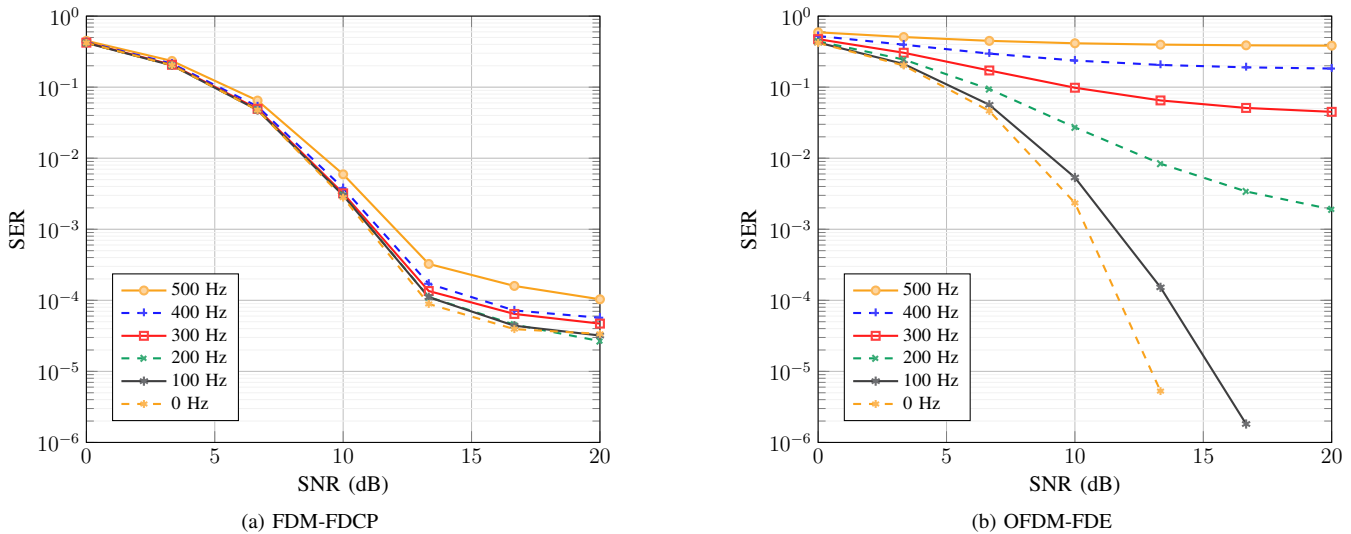

    \centering
    \subfloat[FDM-FDCP]{
        \includestandalone[width=0.45\textwidth]{figures/ser_45_fdm_fdcp}
    }
    \hfill
    \subfloat[OFDM-FDE]{
        \includestandalone[width=0.45\textwidth]{figures/ser_45_ofdm}
    }
    \caption{(a) The decoding performance of 16 QAM using FDM-FDCP with a 4.5 MHz sampling rate and a block length of $N=2\,048$ under AWGN.  The channel model is based on the 3GPP TDL-A model with a normalized delay spread of 20 ns.  In this environment, increasing Doppler spread has little effect on FDM-FDCP whereas (b) OFDM-FDE fails to equalize the channel at high Doppler.}
    \label{fig:awgn2}
\end{figure*}
\fi

In Figure \ref{fig:awgn2} we present an additional set of simulations, where the sampling rate has been increased to 4.5 MHz and the delay spread has been decreased to 20 ns.  
Here, due to the reduced delay spread, we see that the ability of FDM-FDCP to compensate for arbitrary Doppler is only slightly effected by increasing the Doppler spread.  
In contrast, the shorter block length allows OFDM-FDE to more effectively equalize the channel.  
However, once the Doppler spread exceeds 300 Hz, FDM-FDCP outperforms OFDM-FDE.
These results suggest that FDM-FDCP offers an attractive method of transmission in high-Doppler spread channels that are bandlimited either due to constraints imposed by resource allocation or hardware.
At higher sampling rates, our techniques will become more sensitive to delay spread and exhibit a much higher error floor than OFDM-FDE and SC-FDE.

\ifdoublecol
\begin{figure*}
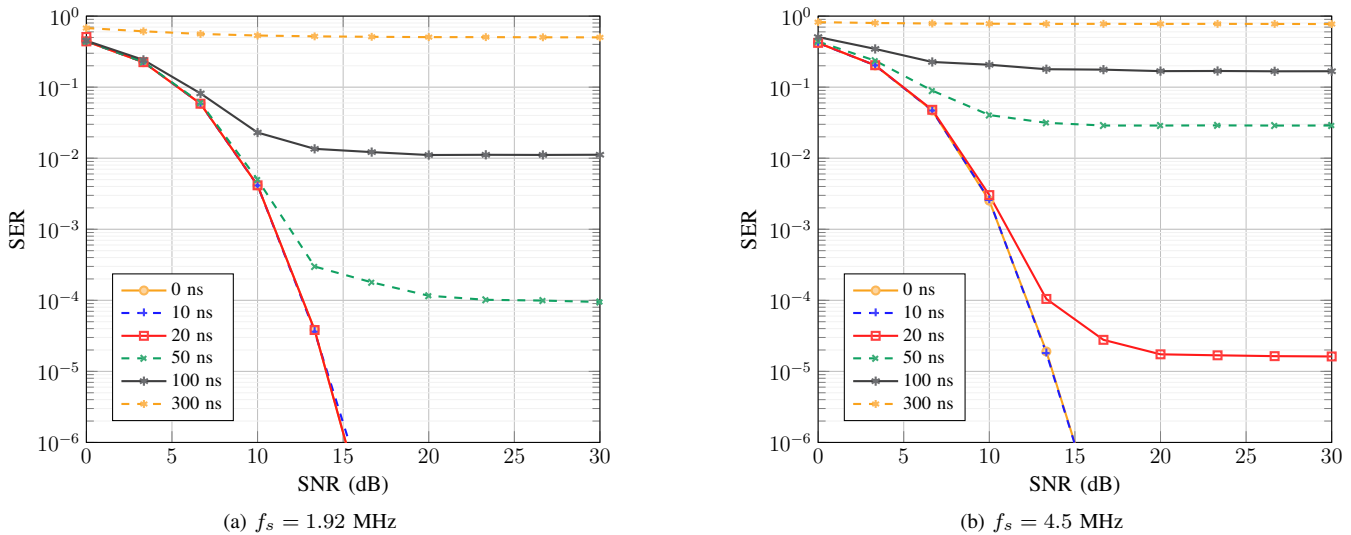

    \centering
    \subfloat[$f_s=1.92$ MHz]{
        \includestandalone[width=0.45\textwidth]{figures/ser_fdm_fdcp_delay_sweep}
    }
    \hfill
    \subfloat[$f_s=4.5$ MHz]{
        \includestandalone[width=0.45\textwidth]{figures/ser_fdm_fdcp_delay_sweep_45}
    }
    \caption{The AWGN performance of FDM-FDCP using 16 QAM for various
delay spreads.  In (a) the sample rate is 1.92 MHz, whereas in (b) the
sample rate is 4.5 MHz. The block length is fixed to $N=2\,048$ and the channel
is drawn from the TDL-A model with a Doppler spread of 200 Hz.  The error floor
remains small so long as the delay spread is small compared to the sample
period.}
    \label{fig:awgn_delay}
\end{figure*}
\fi

\change{
In Figure {\ref{fig:awgn_delay}}, we demonstrate the sensitivity of FDM-FDCP to
the delay spread.  In particular, for these simulations, the Doppler spread is fixed at 200 Hz, while
the delay spread is varied from 0 ns to 300 ns, using the same TDL-A channel
model as in previous simulations.  These simulations show that when the delay
spread is small in comparison to the sample period, the error floor is
negligible.  This performance is analogous to the performance of unequalized single-carrier modulation in channel with negligible Doppler spread and arbitrary delay spread.  
In Section {\ref{sub:parameter_sweep}}, we will explore the error-floor
performance of OFDM-FDE and FDM-FDCP, for a wide variety of sampling rates and delay
spreads, in the absence of AWGN.
}

\subsection{Channel Parameter Sweeps}
\label{sub:channel_sweep}
In this section we characterize the effectiveness of the considered modulation and detection techniques at equalizing channels that are impaired predominantly by a large delay spread or a large Doppler spread.  
We begin by fixing the sampling rate and block length associated with all four modulation and detection techniques and simulate the techniques across a variety of channel conditions.
Specifically, we fix the block length to be $N=1\,024$ and the sampling rate to be 4.5 MHz.  
As with the previous section, the results presented in this section are averaged over $100\, 000$ channel realizations.

In Figure \ref{fig:doppler_sweep}, we fix the delay spread to be 50 ns and sweep the Doppler spread from 50 Hz to $5\,000$ Hz.  
We see that OFDM-FDE and SC-FDE are able to effectively equalize the channel as long as the Doppler period remains small compared to the block duration.  However, for large Doppler spreads, equalization becomes uneffective.
Additionally in Figure \ref{fig:doppler_sweep}, we see that FDM-FDCP and SC-TDE are able to effectively equalize the channel even as the Doppler spread grows. We notice a slight degradation in performance for high Doppler spread channels.  
This is a result of the non-zero delay spread present in the channel; we demonstrate a similar degradation for OFDM-FDE in the next set of simulations.

\ifdoublecol
\begin{figure}
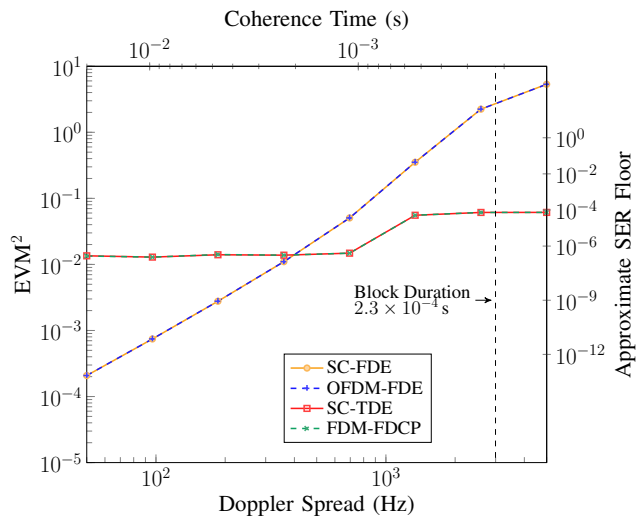

    \includestandalone[width=\figwidth\columnwidth]{evm_plots/doppler_sweep}
    \caption{In this plot, the block length is fixed to $N=1\,024$, the sample rate is fixed to 4.5 MHz, and the delay spread is fixed to 50 ns. As the Doppler spread increases, OFDM-FDE is no longer able to effectively equalize the channel.  Notice that OFDM-FDE becomes ineffective when the duration of the block, here 0.23 ms, is roughly one-tenth of the coherence time of the channel.}
    \label{fig:doppler_sweep}
\end{figure}

\begin{figure}
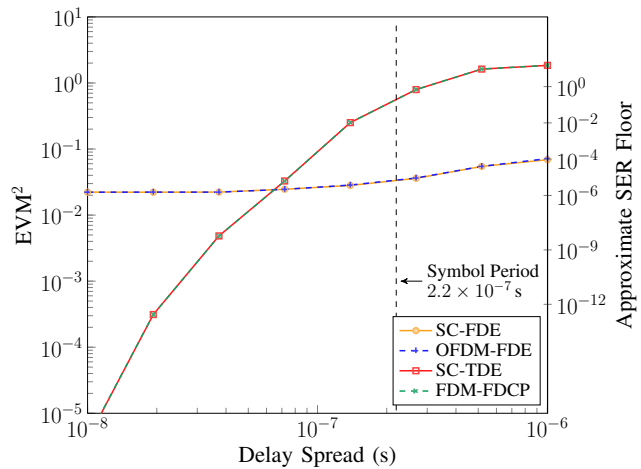

    \vspace{5mm}
    \includestandalone[width=\figwidth\columnwidth]{evm_plots/delay_sweep}
    \caption{Here, the block length is again fixed to $N=1\,024$, the sample rate is fixed to 4.5 MHz.  The Doppler spread is fixed to 500 Hz. As the delay spread increases, our time-domain equalization process is no longer effective.
    Our equalization technique becomes ineffective roughly when the delay spread is close to the symbol period.}
    \label{fig:delay_sweep}
 \end{figure}
\fi
In Figure \ref{fig:delay_sweep}, the Doppler spread is fixed to 500 Hz and the delay spread is swept from 10 ns to $1\,000$ ns.  We observe that FDM-FDCP and SC-TDE are able to effectively equalize these channels as long as the delay spread remains small compared to the symbol period (here $T_s = 22$ ns).  In contrast, the OFDM-FDE and SC-FDE perform well over all delay spreads.  The similarity between Figures \ref{fig:doppler_sweep} and \ref{fig:delay_sweep} should not be surprising as the channel models and modulation and detection techniques can all be related through the principle of time-frequency duality.

\subsection{Waveform Parameter Sweeps}
\label{sub:parameter_sweep}
Modulation and detection techniques based on frequency domain equalization, such as OFDM-FDE and SC-FDE, can effectively compensate for arbitrary delay spread by assuming that the channel is time-invariant over each transmission block.  
The period of the largest Doppler component (approximately the coherence time of the channel) will limit how large $N T_s$ can be while still allowing for effective equalization.  Roughly, these techniques will only effectively equalize the channel if the coherence time of the channel is about ten times larger than $N T_s$. 
In contrast, our modulation and detection techniques, which are based on time-domain equalization, can equalize arbitrary Doppler spread assuming that the delay spread is small in comparison to $T_s$.  As a rule of thumb, we claim that our techniques will effectively equalize Doppler spread if the symbol period is roughly twice the delay spread.

Both of these claims are supported by the simulations shown in Figure \ref{fig:ts_sweep}, which simulates 16-QAM symbols over all four modulation and detection techniques discussed in this paper.
In this set of simulations, we fix the delay spread to be 50 ns and the Doppler spread to be 500 Hz.  The block length is fixed to $N=1\,024$ and the sample rate is varied from 450 kHz to 45 MHz.  
We note that these values extend slightly beyond the range considered by the LTE standard; this is done to emphasize the performance of all techniques at the extremes of high and low sampling rates.
We see that at low sampling rates, FDM-FDCP and SC-TDE substantially outperform other modulation methods.  
However, at high sampling rate OFDM-FDE and SC-FDE outperform these modulation techniques.
Indeed, it is well known that increasing the sampling rate and reducing the transmit block duration is an effective strategy to ensure good performance of OFDM in high-Doppler environments, provided sufficent bandwidth is available.
The results in Figure {\ref{fig:ts_sweep}} support the claim that FDM-FDCP and SC-TDE are attractive methods of transmission in high-Doppler channels that are constrained in bandwidth.
Examples of such channels may include sub-6 GHz high mobility deployments, or mmWave deployments where bandwidth is constrained due to hardware or bandwidth allocation to a very large number of users.

 In Figure \ref{fig:n_sweep}, we present the effects of changing the block length $N$ while holding the sampling rate constant at 4.5 MHz, effectively shortening the transmit block duration ($N T_s$). These simulations use the same parameters as those simulations presented in Figure \ref{fig:ts_sweep}.  As expected, the error floor for OFDM-FDE vanishes for small block lengths, whereas the error floor for FDM-FDCP and SC-TDE is not effected by varying the block length.

Finally, we note that in both Figures {\ref{fig:ts_sweep}} and {\ref{fig:n_sweep}}, the error floor performance of FDM-FDCP and SC-TDE as well as that of OFDM-FDE and SC-FDE are identical.
Since FDM-FDCP offers less complex modulation and demodulation than SC-TDE, this suggests that, if CSI is not known at the transmitter, it is likely advantageous to use FDM-FDCP.
However, it is well known that the use of adaptive loading with OFDM can offer substantial performance improvements over ordinary OFDM-FDE or SC-FDE (see, for example, \cite{rhee2000increase}).
Since adaptive loading is possible with SC-TDE and not FDM-FDCP, it is likely that the use of SC-TDE offers a higher maximum achievable rate if transmit CSI is known and can be adapted to.
Adaptive loading in SC-TDE is thus an interesting topic for research.

\ifdoublecol
\begin{figure}
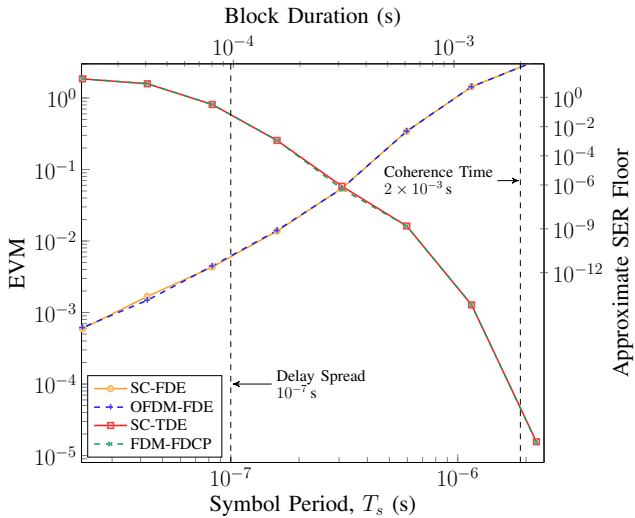

    \includestandalone[width=\figwidth\columnwidth]{evm_plots/symbol_rate_sweep}
    \caption{Here the block length is fixed to $N=1\,024$ and the sampling rate is swept from 450 kHz to 45 MHz.  The channel has a delay spread of 100 ns and a Doppler spread of 500 Hz.  Our techniques perform well as long as the symbol period is small, or roughly the same order, compared to the delay spread.  In contrast, OFDM-FDE and SC-FDE perform well if the block length is much smaller than the coherence time of the channel.}
    \label{fig:ts_sweep}
\end{figure}
\begin{figure}
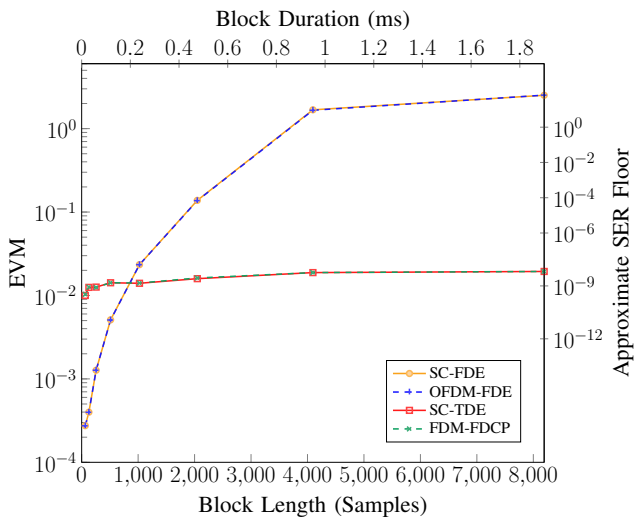

        \includestandalone[width=\figwidth\columnwidth]{evm_plots/block_length_sweep}
    \caption{Here, the sample rate is fixed to 4.5 MHz and the block length is swept from $N=64$ to $N=8\,192$.  The delay spread is 50 ns and the Doppler spread is 500 Hz.  Changing the block length has little effect on our modulation and detection techniques.  However, frequency-domain equalization becomes ineffective if the duration of the block becomes roughly one-fifth the coherence time.}
    \label{fig:n_sweep}
\end{figure}
\fi

\section{Conclusions}
\label{sec:conclusion}

Many modulation and detection techniques used in current wireless systems, such as OFDM-FDE or SC-FDE, make use of the assumption of time-invariance, or approximate time-invariance, of the channel in order to efficiently equalize effects of inter-symbol interference.  However, in practice, time variations in channel responses may occur at time scales much smaller that the duration of the transmission block. 
These time variations may occur in wireless channels associated with mobile transceivers, high carrier frequencies, scheduling/resource allocation time scales, or large noise floors that necessitate time-averaging for noise suppression. As demonstrated in this paper, OFDM-FDE is no longer competitive for such time-varying channels.  For such environments, we propose two new modulation and detection techniques, FDM-FDCP and SC-TDE, both of which use a frequency-domain cyclic prefix.  
\change{
Disregarding the frequency-domain cyclic prefix, these proposed waveforms are modulated
in an identical manner to windowed versions of SC-FDE and OFDM-FDE, but with
different methods for channel estimation and equalization. 
Moreover, the inclusion of the FDCP converts the time-varying channel to a static
channel via a circular convolution, analogous to the use of the time-domain cyclic prefix used in OFDM-FDE.
This allows us to leverage knowledge and intuition developed from studying OFDM to design waveforms for
high-Doppler channels.
For example, one may apply many of the same principles used to choose waveform parameters and windowing functions, or design receiver architectures, to the deployment of FDM-FDCP and SC-TDE.
}

A complete evaluation of the relative benefits of different modulation and detection techniques in wireless channels would depend critically on the joint delay and Doppler spreads in the wireless channel.  In this work, we show that that FDM-FDCP and SC-TDE can outperform OFDM-FDE and SC-FDE in situations where there is a significant Doppler spread and a low-to-moderate delay spread.  
In order to effectively mitigate the effects of Doppler spread, FDM-FDCP and
SC-TDE require an overhead in bandwidth that is proportional to the Doppler
spread of the channel.  As described in this paper, the modulation and detection
of these techniques can be implemented with a time complexity of $O(N \log N)$,
making them competitive with both OFDM-FDE and SC-FDE in runtime and power
consumption.   

We do not explore the effect of a variety of design parameters such as choice of pulse shape, or channel estimation algorithms on the performance of our waveforms.  
We leave such comparisons as a topic of future work. 
Further, our work suggests that the error-floor performance of SC-TDE and FDM-FDCP is approximately identical.  
Since FDM-FDCP affords less complex modulation and demodulation that SC-TDE, its use may be advantageous if CSI is not available at the transmitter.  
However, if transmit CSI is available, it is possible to apply adaptive loading to the SC-TDE.  
It is an open question to explore the performance of SC-TDE under adaptive loading and its relationship to information-theoretic bounds of performance in channels with high-Doppler and low-delay spread.
Finally, this work focused on only one-dimensional waveforms that require $O(N \log N)$ equalization.
As such, a quantitative comparison with techniques with more complex equalization that are designed for general time-varying channels, such as two-dimensional signaling or time-frequency modulation, are also left as a topic of future research.

\section*{Acknowledgements}
The authors would like to thank Ronny Hadani and Anton Monk for discussions on the OTFS waveform and time-frequency representations of signals, Ali Khayrallah for his comments on a preliminary version of this work, and the anonymous reviewers for their extremely constructive and comprehensive comments on this work.
\ifdoublecol
\balance
\fi

\bibliographystyle{IEEEtran}
\bibliography{IEEEabrv,references}
\end{document}

%% file: figures/fdcp_tikz.tex
\begin{tikzpicture}[
    every node/.style={scale=0.75},
    bin/.style={top color=red!30, bottom color=red!10, draw=red!60}
]
\foreach \x in {0.0, 0.5, ..., 6}
{\filldraw[fill=none, draw=blue!75] (\x,0) rectangle (\x+0.5,2);}

\foreach \x/\y in {3.5/60, 4/60, 4.5/60, 5/60, 5.5/60} {
\filldraw[bin] (\x+0.5,0) rectangle (\x+1,\y/30);
}

\foreach \x/\y in {3.5/60, 4/60, 4.5/60, 5/60, 5.5/60} {
\filldraw[bin] (\x+0.5-4,0) rectangle (\x+1-4,\y/30);
}

\draw [very thick, ->] (5.25, 2.1) to [out=135,in=45] (1.25, 2.1);
\draw[->] (0,0) -- (0, 3);
\draw[->] (0,0) -- (7, 0);
\node at (3.5, -0.85) {Subcarrier index};
\node[rotate=90] at (-0.5, 1.5) {Amplitude};
\node at (4.75+0.5, -0.3) {\large $s$};
\node at (5.75+0.8, -0.3) {\large $s+f_\text{max}$};
\node at (3.75+0.5, -0.3) {\large $s-f_\text{max}$};
\end{tikzpicture}

%% file: figures/zero.tex
\begin{tikzpicture}
\node at (6.8, -0.4) {Frequency};
\node[rotate=90] at (-0.6, 1.5) {Amplitude};

\shadedraw[top color=black!20, bottom color=white] (2,0) rectangle (5,2);

\coordinate (t0) at (1,0);
\coordinate (t1) at (2,0);
\coordinate (t2) at (2,2);

\coordinate (s0) at (5,0);
\coordinate (s1) at (6,0);
\coordinate (s2) at (5,2);

\coordinate (v1) at (3,0);
\coordinate (v2) at (4,0);

\shadedraw[top color=red!50, bottom color=red!10, draw=red] (t0) -- (t1) -- (t2) [snake=saw] -- cycle;
\shadedraw[top color=blue!50, bottom color=blue!10, draw=blue!50] (s1) -- (s0) -- (s2) [snake=saw] -- cycle;
\shadedraw[top color=blue!50, bottom color=blue!10, draw=blue!50]  (t1) -- (t2) [snake=saw] -- (v1) [snake=none] -- cycle;
\shadedraw[top color=red!50, bottom color=red!10, draw=red] (v2) -- (s0) -- (s2) [snake=saw] --  cycle;

\draw[ss] (5.5,0) |- (2.5,-0.5) -- (2.5,0);
\draw[ss] (1.8,1.7) |- (4.8,2.25) -- (4.8,1.7);

\draw[black] (s0)--(s2)--(t2)--(t1);
\node[align=center,font=\large] at (3.5,1) 
{\textbf{FDM-FDCP}\\\textbf{Symbols}};

\draw[->] (0,0) -- (0, 3);
\draw[->] (0,0) -- (7, 0);
\end{tikzpicture}

%% file: figures/channel_estimation.tex
\begin{tikzpicture}

\node at (7.5, -0.4) {Frequency};
\node[rotate=90] at (-0.6, 1.5) {Amplitude};

\coordinate (t0) at (0,0);
\coordinate (t1) at (0,0);
\coordinate (t2) at (0,2);

\coordinate (s0) at (2,0);
\coordinate (s1) at (2.9,0);
\coordinate (s2) at (2,2);

\coordinate (t02) at (6,2);
\coordinate (t12) at (6,0);
\coordinate (t22) at (5.1,0);

\coordinate (s02) at (8,0);
\coordinate (s12) at (8,0);
\coordinate (s22) at (8,2);

\shadedraw[top color=black!20, bottom color=white] (t1) rectangle (s2);

\shadedraw[top color=black!20, bottom color=white] (t12) rectangle (s22);

\shadedraw[top color=red!50, bottom color=red!10, draw=red!50] (s1) -- (s0) -- (2,1.75) .. controls(2.25,0.1) and (2.75,0.1) .. cycle;

\shadedraw[top color=red!50, bottom color=red!10, draw=red!50] (t22) -- (t12) -- (6,1.75) .. controls(5.75,0.1) and (5.25,0.1) .. cycle;

\draw[black] (s0)--(s2)--(t2)--(t1);
\draw[black] (s02)--(s22)--(t02)--(t12);

\node[align=center] at ($(s2)!0.5!(t1)$) 
{\textbf{Data}\\\textbf{Symbols}};
\node[align=center] at ($(s22)!0.5!(t12)$) 
{\textbf{Data}\\\textbf{Symbols}};

\shadedraw[top color=blue!50, bottom color=blue!10, draw=blue!50] (3.1,0) -- (4,0) -- (4,1.75) .. controls(3.75,0.1) and (3.25,0.1) .. cycle;
\shadedraw[top color=blue!50, bottom color=blue!10, draw=blue!50] (4.9,0) -- (4,0) -- (4,1.75) .. controls(4.25,0.1) and (4.75,0.1) .. (5,0);

\draw[thick,-stealth] (4,0) -- (4,2) node[above] {Pilot};

\draw[->] (0,0) -- (0, 3);
\draw[->] (0,0) -- (8.5, 0);

\draw[decorate,decoration=brace] (5,-0.05) -- (3,-0.05) node[below, midway] {$\hat{H}(f)$};
\end{tikzpicture}

%% file: figures/system_model.tex
\begin{tikzpicture}[
    every node/.style={
        draw,
        align=center,
        minimum size=14mm,
        top color=Blue!20, 
        bottom color=white
    }
]

\node[top color=Black!20] (data) {Data \\ Symbols};
\node[right = of data] (s2p) {Serial to \\ Parallel};
\node[right = of s2p] (pilot) {Pilot \\ Insertion};
\node[right = of pilot] (FDCP) {FDCP \\ Insertion};
\node[right = of FDCP] (finv) {$\mathcal{F}^{-1}$};
\node[right = of finv] (pulse) {Pulse \\ Shaping};
\node[right = of pulse] (da) {D/A};

\node[below = of da] (ad) {A/D};
\node[below = of pulse] (equal) {Equalize};
\node[below = of finv] (f2) {$\mathcal{F}$};
\node[below = of FDCP] (FDCPinv) {FDCP \\ Removal};
\node[below = of pilot] (pilotinv) {Pilot \\ Removal};
\node[below = of s2p] (p2s) {Parallel \\ to Serial};
\node[top color=Black!20,below = of data] (data2) {Data \\ Symbols};

\node[top color=Red!20] at ($(da.east)!0.5!(ad.east) + (1.5,0)$) (channel) {$h(t)$};

\node[top color=Black!20] at ($(equal) - (1.2,2.2)$) (CSI) {CSI};
\node[top color=Black!20] at ($(equal) - (-1.2,2.2)$) (pulse2) {Pulse \\ Shape};

\draw[->, thick] (data) -- (s2p);
\draw[->, thick] (s2p) -- (pilot);
\draw[->, thick] (pilot) -- (FDCP);
\draw[->, thick] (FDCP) -- (finv);
\draw[->, thick] (finv) -- (pulse);
\draw[->, thick] (pulse) -- (da);
\draw[->, thick] (da) -| (channel);

\draw[->, thick] (channel) |- (ad);
\draw[->, thick] (ad) -- (equal);
\draw[->, thick] (equal) -- (f2);
\draw[->, thick] (f2) -- (FDCPinv);
\draw[->, thick] (FDCPinv) -- (pilotinv);
\draw[->, thick] (pilotinv) -- (p2s);
\draw[->, thick] (p2s) -- (data2);

\draw[->, thick] (CSI.north) -- (equal);
\draw[->, thick] (pulse2.north) -- (equal);


\end{tikzpicture}

%% file: figures/system_sc_tde.tex
\begin{tikzpicture}[
    every node/.style={
        draw,
        align=center,
        minimum size=14mm,
        top color=Blue!20, 
        bottom color=white
    }
]

\node[top color=Black!20] (data) {Data \\ Symbols};
\node[right = of data] (s2p) {Serial to \\ Parallel};
\node[right = of s2p] (pilot) {Pilot \\ Insertion};
\node[right = of pilot] (f) {$\mathcal{F}$};
\node[right = of f] (FDCP) {FDCP \\ Insertion};
\node[right = of FDCP] (finv) {Pulse \\ Shaping};
\node[right = of finv] (pulse) {$\mathcal{F}^{-1}$};
\node[right = of pulse] (da) {D/A};

\node[below = of da] (ad) {A/D};
\node[below = of pulse] (equal) {Equalize};
\node[below = of finv] (f2) {$\mathcal{F}$};
\node[below = of FDCP] (FDCPinv) {FDCP \\ Removal};
\node[below = of f] (finv2) {$\mathcal{F}^{-1}$};
\node[below = of pilot] (pilotinv) {Pilot \\ Removal};
\node[below = of s2p] (p2s) {Parallel \\ to Serial};
\node[top color=Black!20,below = of data] (data2) {Data \\ Symbols};

\node[top color=Red!20] at ($(da.east)!0.5!(ad.east) + (1.5,0)$) (channel) {$h(t)$};

\node[top color=Black!20] at ($(equal) - (1.2,2.2)$) (CSI) {CSI};
\node[top color=Black!20] at ($(equal) - (-1.2,2.2)$) (pulse2) {Pulse \\ Shape};

\draw[->, thick] (data) -- (s2p);
\draw[->, thick] (s2p) -- (pilot);
\draw[->, thick] (pilot) -- (f);
\draw[->, thick] (f) -- (FDCP);
\draw[->, thick] (FDCP) -- (finv);
\draw[->, thick] (finv) -- (pulse);
\draw[->, thick] (pulse) -- (da);
\draw[->, thick] (da) -| (channel);

\draw[->, thick] (channel) |- (ad);
\draw[->, thick] (ad) -- (equal);
\draw[->, thick] (equal) -- (f2);
\draw[->, thick] (f2) -- (FDCPinv);
\draw[->, thick] (FDCPinv) -- (finv2);
\draw[->, thick] (finv2) -- (pilotinv);
\draw[->, thick] (pilotinv) -- (p2s);
\draw[->, thick] (p2s) -- (data2);

\draw[->, thick] (CSI.north) -- (equal);
\draw[->, thick] (pulse2.north) -- (equal);

\end{tikzpicture}

%% file: figures/fading.tex
\begin{tikzpicture}[every node/.style={scale=0.70}]
\begin{axis}[
    xmin=0, xmax=8000,
    ymin=0,
	ylabel={\Large $|h[n]|$},
    xlabel={\Large $n$ (samples)},
    tick label style={font=\large}
]

\addplot[very thick, red, mark=none, smooth] 
coordinates
{
(1,8.6212e-01)
(129,5.3849e-01)
(257,3.2968e-01)
(385,5.3249e-01)
(513,7.4441e-01)
(641,9.5723e-01)
(769,1.1551)
(897,1.3274)
(1025,1.4671)
(1153,1.5693)
(1281,1.6307)
(1409,1.6493)
(1537,1.6245)
(1665,1.5572)
(1793,1.4494)
(1921,1.3048)
(2049,1.1282)
(2177,9.2661e-01)
(2305,7.1018e-01)
(2433,4.9798e-01)
(2561,3.4123e-01)
(2689,3.5057e-01)
(2817,5.1510e-01)
(2945,7.2581e-01)
(3073,9.3670e-01)
(3201,1.1303)
(3329,1.2971)
(3457,1.4307)
(3585,1.5266)
(3713,1.5818)
(3841,1.5945)
(3969,1.5643)
(4097,1.4924)
(4225,1.3809)
(4353,1.2333)
(4481,1.0547)
(4609,8.5159e-01)
(4737,6.3376e-01)
(4865,4.1998e-01)
(4993,2.7032e-01)
(5121,3.1900e-01)
(5249,5.0776e-01)
(5377,7.2288e-01)
(5505,9.3021e-01)
(5633,1.1166)
(5761,1.2744)
(5889,1.3980)
(6017,1.4835)
(6145,1.5282)
(6273,1.5309)
(6401,1.4914)
(6529,1.4111)
(6657,1.2924)
(6785,1.1391)
(6913,9.5613e-01)
(7041,7.5029e-01)
(7169,5.3160e-01)
(7297,3.2244e-01)
(7425,2.1538e-01)
(7553,3.4000e-01)
(7681,5.5060e-01)
(7809,7.6711e-01)
(7937,9.5950e-01)
(8065,1.0164)
};

\addplot[very thick, blue, dashed, mark=none,smooth]
coordinates
{
(1,7.8754e-01)
(129,6.6861e-01)
(257,7.3067e-01)
(385,8.6195e-01)
(513,9.3970e-01)
(641,1.0303)
(769,1.1246)
(897,1.2124)
(1025,1.2862)
(1153,1.3406)
(1281,1.3724)
(1409,1.3795)
(1537,1.3616)
(1665,1.3197)
(1793,1.2565)
(1921,1.1762)
(2049,1.0851)
(2177,9.9165e-01)
(2305,9.0717e-01)
(2433,8.4494e-01)
(2561,8.1747e-01)
(2689,8.3123e-01)
(2817,8.8280e-01)
(2945,9.6098e-01)
(3073,1.0523)
(3201,1.1447)
(3329,1.2288)
(3457,1.2976)
(3585,1.3464)
(3713,1.3721)
(3841,1.3732)
(3969,1.3497)
(4097,1.3030)
(4225,1.2361)
(4353,1.1537)
(4481,1.0626)
(4609,9.7190e-01)
(4737,8.9326e-01)
(4865,8.3984e-01)
(4993,8.2279e-01)
(5121,8.4609e-01)
(5249,9.0416e-01)
(5377,9.8508e-01)
(5505,1.0758)
(5633,1.1651)
(5761,1.2443)
(5889,1.3071)
(6017,1.3493)
(6145,1.3681)
(6273,1.3625)
(6401,1.3329)
(6529,1.2811)
(6657,1.2106)
(6785,1.1266)
(6913,1.0364)
(7041,9.4983e-01)
(7169,8.7891e-01)
(7297,8.3633e-01)
(7425,8.3136e-01)
(7553,8.6510e-01)
(7681,9.2988e-01)
(7809,1.0115)
(7937,1.0341)
(8065,9.5072e-01)
};

\addplot[black, very thick, mark=none, smooth, densely dotted]
coordinates
{
(1,8.9845e-01)
(129,4.7976e-01)
(257,2.2681e-04)
(385,4.3158e-01)
(513,6.9022e-01)
(641,9.2448e-01)
(769,1.1328)
(897,1.3092)
(1025,1.4489)
(1153,1.5478)
(1281,1.6033)
(1409,1.6138)
(1537,1.5789)
(1665,1.4996)
(1793,1.3783)
(1921,1.2182)
(2049,1.0238)
(2177,8.0075e-01)
(2305,5.5516e-01)
(2433,2.9401e-01)
(2561,2.5991e-04)
(2689,2.4601e-01)
(2817,5.0926e-01)
(2945,7.5830e-01)
(3073,9.8610e-01)
(3201,1.1862)
(3329,1.3531)
(3457,1.4820)
(3585,1.5694)
(3713,1.6127)
(3841,1.6108)
(3969,1.5636)
(4097,1.4727)
(4225,1.3404)
(4353,1.1705)
(4481,9.6773e-01)
(4609,7.3788e-01)
(4737,4.8735e-01)
(4865,2.2325e-01)
(4993,4.8198e-04)
(5121,3.1656e-01)
(5249,5.7665e-01)
(5377,8.2057e-01)
(5505,1.0414)
(5633,1.2330)
(5761,1.3900)
(5889,1.5079)
(6017,1.5834)
(6145,1.6145)
(6273,1.6002)
(6401,1.5409)
(6529,1.4384)
(6657,1.2954)
(6785,1.1161)
(6913,9.0539e-01)
(7041,6.6925e-01)
(7169,4.1432e-01)
(7297,1.4787e-01)
(7425,1.2350e-01)
(7553,3.9064e-01)
(7681,6.4702e-01)
(7809,8.8576e-01)
(7937,1.0792)
(8065,1.1083)};

\end{axis}
\end{tikzpicture}

%% file: figures/ser_fdm_fdcp.tex
\begin{tikzpicture}[every node/.style={scale=0.70}]
\begin{semilogyaxis}[
    xmin=0,xmax=30,
	grid=both,
	grid style={line width=.1pt, draw=gray!10},
	major grid style={line width=.2pt,draw=gray!50},
    ymin=1e-6, ymax=1,
    cycle list name=set1,
	ylabel={SER},
    xlabel={SNR (dB)},
    label style={font=\large},
    tick label style={font=\large}, 
    legend style={font=\normalsize,
    			  cells={anchor=west},
    			  anchor=south west,
                  at={(0.05,0.05)}}
]

\addplot 
coordinates
{
(0.000, 5.641e-01)
(3.333, 3.728e-01)
(6.667, 1.828e-01)
(10.000, 6.736e-02)
(13.333, 2.758e-02)
(16.667, 1.416e-02)
(30.000, 2.527e-03)
};

\addplot 
coordinates
{
(0.000, 5.169e-01)
(3.333, 3.190e-01)
(6.667, 1.409e-01)
(10.000, 5.120e-02)
(13.333, 2.249e-02)
(16.667, 1.189e-02)
(30.000, 2.717e-03)
};

\addplot 
coordinates
{
(0.000, 5.086e-01)
(3.333, 3.101e-01)
(6.667, 1.402e-01)
(10.000, 5.534e-02)
(13.333, 2.659e-02)
(16.667, 1.574e-02)
(30.000, 3.577e-03)
};

\addplot 
coordinates
{
(0.000, 4.423e-01)
(3.333, 2.261e-01)
(6.667, 6.020e-02)
(10.000, 5.110e-03)
(13.333, 3.475e-04)
(16.667, 1.593e-04)
(30.000, 9.789e-05)
};

\addplot 
coordinates
{
(0.000, 4.210e-01)
(3.333, 2.044e-01)
(6.667, 4.870e-02)
(10.000, 3.173e-03)
(13.333, 1.615e-04)
(16.667, 9.139e-05)
(30.000, 4.413e-05)
};

\addplot 
coordinates
{
(0.000, 4.202e-01)
(3.333, 2.035e-01)
(6.667, 4.793e-02)
(10.000, 2.940e-03)
(13.333, 1.543e-04)
(16.667, 7.869e-05)
(30.000, 4.810e-05)
};

\legend{500 Hz,400 Hz,300 Hz,200 Hz,100 Hz, 0 Hz}
\end{semilogyaxis}
\end{tikzpicture}

%% file: figures/ser_ofdm.tex
\begin{tikzpicture}[every node/.style={scale=0.70}]
\begin{semilogyaxis}[
    xmin=0,xmax=30,
	grid=both,
	grid style={line width=.1pt, draw=gray!10},
	major grid style={line width=.2pt,draw=gray!50},
    ymin=1e-6, ymax=1,
    cycle list name=set1,
	ylabel={SER},
    xlabel={SNR (dB)},
    label style={font=\large},
    tick label style={font=\large},
    legend style={font=\normalsize,
    			  cells={anchor=west},
    			  anchor=south west,
                  at={(0.05,0.05)}}
]

\addplot 
coordinates
{
(0.000, 8.695e-01)
(3.333, 8.655e-01)
(6.667, 8.641e-01)
(10.000, 8.630e-01)
(13.333, 8.625e-01)
(16.667, 8.629e-01)
(30.000, 8.620e-01)
};

\addplot 
coordinates
{
(0.000, 8.268e-01)
(3.333, 8.183e-01)
(6.667, 8.147e-01)
(10.000, 8.119e-01)
(13.333, 8.101e-01)
(16.667, 8.097e-01)
(30.000, 8.104e-01)
};

\addplot 
coordinates
{
(0.000, 7.299e-01)
(3.333, 7.022e-01)
(6.667, 6.870e-01)
(10.000, 6.790e-01)
(13.333, 6.750e-01)
(16.667, 6.732e-01)
(30.000, 6.716e-01)
};

\addplot 
coordinates
{
(0.000, 5.690e-01)
(3.333, 4.715e-01)
(6.667, 4.011e-01)
(10.000, 3.593e-01)
(13.333, 3.373e-01)
(16.667, 3.265e-01)
(30.000, 3.175e-01)
};

\addplot 
coordinates
{
(0.000, 4.510e-01)
(3.333, 2.622e-01)
(6.667, 1.156e-01)
(10.000, 4.436e-02)
(13.333, 1.987e-02)
(16.667, 1.149e-02)
(30.000, 6.222e-03)
};

\addplot 
coordinates
{
(0.000, 4.193e-01)
(3.333, 2.019e-01)
(6.667, 4.631e-02)
(10.000, 2.351e-03)
(13.333, 5.589e-06)
(16.667, 0.000e+00)
(30.000, 0.000e+00)
};

\legend{500 Hz,400 Hz,300 Hz,200 Hz,100 Hz, 0 Hz}
\end{semilogyaxis}
\end{tikzpicture}

%% file: figures/ser_45_fdm_fdcp.tex
\begin{tikzpicture}[every node/.style={scale=0.70}]
\begin{semilogyaxis}[
    xmin=0,xmax=20,
	grid=both,
	grid style={line width=.1pt, draw=gray!10},
	major grid style={line width=.2pt,draw=gray!50},
    ymin=1e-6, ymax=1,
    cycle list name=set1,
	ylabel={SER},
    xlabel={SNR (dB)},
    label style={font=\large},
    tick label style={font=\large},
    legend style={font=\normalsize,
    			  cells={anchor=west},
    			  anchor=south west,
                  at={(0.05,0.05)}}
]

\addplot 
coordinates
{
(0.000, 4.493e-01)
(3.333, 2.339e-01)
(6.667, 6.476e-02)
(10.000, 5.936e-03)
(13.333, 3.253e-04)
(16.667, 1.594e-04)
(20.000, 1.034e-04)
};

\addplot 
coordinates
{
(0.000, 4.312e-01)
(3.333, 2.144e-01)
(6.667, 5.355e-02)
(10.000, 3.789e-03)
(13.333, 1.713e-04)
(16.667, 7.200e-05)
(20.000, 5.686e-05)
};

\addplot 
coordinates
{
(0.000, 4.231e-01)
(3.333, 2.064e-01)
(6.667, 4.946e-02)
(10.000, 3.202e-03)
(13.333, 1.352e-04)
(16.667, 6.432e-05)
(20.000, 4.702e-05)
};

\addplot 
coordinates
{
(0.000, 4.201e-01)
(3.333, 2.035e-01)
(6.667, 4.807e-02)
(10.000, 3.026e-03)
(13.333, 1.112e-04)
(16.667, 4.549e-05)
(20.000, 2.668e-05)
};

\addplot 
coordinates
{
(0.000, 4.193e-01)
(3.333, 2.026e-01)
(6.667, 4.763e-02)
(10.000, 2.993e-03)
(13.333, 1.106e-04)
(16.667, 4.392e-05)
(20.000, 3.197e-05)
};

\addplot 
coordinates
{
(0.000, 4.200e-01)
(3.333, 2.031e-01)
(6.667, 4.744e-02)
(10.000, 2.791e-03)
(13.333, 8.885e-05)
(16.667, 3.934e-05)
(20.000, 3.373e-05)
};

\legend{500 Hz,400 Hz,300 Hz,200 Hz,100 Hz, 0 Hz}
\end{semilogyaxis}
\end{tikzpicture}

%% file: figures/ser_45_ofdm.tex
\begin{tikzpicture}[every node/.style={scale=0.70}]
\begin{semilogyaxis}[
    xmin=0,xmax=20,
	grid=both,
	grid style={line width=.1pt, draw=gray!10},
	major grid style={line width=.2pt,draw=gray!50},
    ymin=1e-6, ymax=1,
    cycle list name=set1,
	ylabel={SER},
    xlabel={SNR (dB)},
    label style={font=\large},
    tick label style={font=\large},
    legend style={font=\normalsize,
    			  cells={anchor=west},
    			  anchor=south west,
                  at={(0.05,0.05)}}
]

\addplot 
coordinates
{
(0.000, 5.912e-01)
(3.333, 5.074e-01)
(6.667, 4.485e-01)
(10.000, 4.145e-01)
(13.333, 3.969e-01)
(16.667, 3.881e-01)
(20.000, 3.839e-01)
};

\addplot 
coordinates
{
(0.000, 5.246e-01)
(3.333, 3.959e-01)
(6.667, 2.967e-01)
(10.000, 2.368e-01)
(13.333, 2.054e-01)
(16.667, 1.901e-01)
(20.000, 1.827e-01)
};

\addplot 
coordinates
{
(0.000, 4.741e-01)
(3.333, 3.049e-01)
(6.667, 1.717e-01)
(10.000, 9.816e-02)
(13.333, 6.506e-02)
(16.667, 5.100e-02)
(20.000, 4.494e-02)
};

\addplot 
coordinates
{
(0.000, 4.420e-01)
(3.333, 2.449e-01)
(6.667, 9.389e-02)
(10.000, 2.718e-02)
(13.333, 8.374e-03)
(16.667, 3.405e-03)
(20.000, 1.906e-03)
};

\addplot 
coordinates
{
(0.000, 4.247e-01)
(3.333, 2.120e-01)
(6.667, 5.648e-02)
(10.000, 5.352e-03)
(13.333, 1.523e-04)
(16.667, 1.817e-06)
};

\addplot 
coordinates
{
(0.000, 4.193e-01)
(3.333, 2.019e-01)
(6.667, 4.625e-02)
(10.000, 2.346e-03)
(13.333, 5.279e-06)
(16.667, 0.000e+00)
(20.000, 0.000e+00)
};

\legend{500 Hz,400 Hz,300 Hz,200 Hz,100 Hz, 0 Hz}
\end{semilogyaxis}
\end{tikzpicture}

%% file: figures/ser_fdm_fdcp_delay_sweep.tex
\begin{tikzpicture}[every node/.style={scale=0.70}]
\begin{semilogyaxis}[
    xmin=0,xmax=30,
	grid=both,
	grid style={line width=.1pt, draw=gray!10},
	major grid style={line width=.2pt,draw=gray!50},
    ymin=1e-6, ymax=1,
    cycle list name=set1,
	ylabel={SER},
    xlabel={SNR (dB)},
    label style={font=\large},
    tick label style={font=\large},
    legend style={font=\normalsize,
    			  cells={anchor=west},
    			  anchor=south west,
                  at={(0.05,0.05)}}
]

\addplot 
coordinates
{
(0.000, 4.414e-01)
(3.333, 2.246e-01)
(6.667, 5.843e-02)
(10.000, 4.150e-03)
(13.333, 3.845e-05)
(16.667, 4.883e-08)
};

\addplot 
coordinates
{
(0.000, 4.414e-01)
(3.333, 2.250e-01)
(6.667, 5.843e-02)
(10.000, 4.132e-03)
(13.333, 3.621e-05)
(16.667, 9.766e-08)
};

\addplot 
coordinates
{
(0.000, 5.086e-01)
(0.000, 4.412e-01)
(3.333, 2.248e-01)
(6.667, 5.850e-02)
(10.000, 4.155e-03)
(13.333, 3.821e-05)
(16.667, 4.883e-08)
};

\addplot 
coordinates
{
(0.000, 4.423e-01)
(3.333, 2.261e-01)
(6.667, 6.041e-02)
(10.000, 5.001e-03)
(13.333, 2.992e-04)
(16.667, 1.793e-04)
(20.000, 1.160e-04)
(23.333, 1.017e-04)
(26.667, 9.919e-05)
(30.000, 9.475e-05)
};

\addplot 
coordinates
{
(0.000, 4.504e-01)
(3.333, 2.420e-01)
(6.667, 8.164e-02)
(10.000, 2.297e-02)
(13.333, 1.355e-02)
(16.667, 1.219e-02)
(20.000, 1.109e-02)
(23.333, 1.117e-02)
(26.667, 1.112e-02)
(30.000, 1.119e-02)
};

\addplot 
coordinates
{
(0.000, 6.787e-01)
(3.333, 6.111e-01)
(6.667, 5.608e-01)
(10.000, 5.330e-01)
(13.333, 5.188e-01)
(16.667, 5.113e-01)
(20.000, 5.059e-01)
(23.333, 5.060e-01)
(26.667, 5.032e-01)
(30.000, 5.019e-01)
};

\legend{0 ns,10 ns,20 ns,50 ns,100 ns, 300 ns}
\end{semilogyaxis}
\end{tikzpicture}

%% file: figures/ser_fdm_fdcp_delay_sweep_45.tex
\begin{tikzpicture}[every node/.style={scale=0.70}]
\begin{semilogyaxis}[
    xmin=0,xmax=30,
	grid=both,
	grid style={line width=.1pt, draw=gray!10},
	major grid style={line width=.2pt,draw=gray!50},
    ymin=1e-6, ymax=1,
    cycle list name=set1,
	ylabel={SER},
    xlabel={SNR (dB)},
    label style={font=\large},
    tick label style={font=\large},
    legend style={font=\normalsize,
    			  cells={anchor=west},
    			  anchor=south west,
                  at={(0.05,0.05)}}
]

\addplot 
coordinates
{
(0.000, 4.195e-01)
(3.333, 2.023e-01)
(6.667, 4.682e-02)
(10.000, 2.529e-03)
(13.333, 1.904e-05)
(16.667, 4.883e-08)
};

\addplot 
coordinates
{
(0.000, 4.190e-01)
(3.333, 2.025e-01)
(6.667, 4.691e-02)
(10.000, 2.583e-03)
(13.333, 1.807e-05)
(16.667, 4.883e-08)
};

\addplot 
coordinates
{
(0.000, 4.201e-01)
(3.333, 2.036e-01)
(6.667, 4.802e-02)
(10.000, 3.028e-03)
(13.333, 1.046e-04)
(16.667, 2.778e-05)
(20.000, 1.738e-05)
(23.333, 1.683e-05)
(26.667, 1.638e-05)
(30.000, 1.622e-05)

};

\addplot 
coordinates
{
(0.000, 4.388e-01)
(3.333, 2.356e-01)
(6.667, 8.943e-02)
(10.000, 4.050e-02)
(13.333, 3.149e-02)
(16.667, 2.884e-02)
(20.000, 2.883e-02)
(23.333, 2.897e-02)
(26.667, 2.881e-02)
(30.000, 2.885e-02)
};

\addplot 
coordinates
{
(0.000, 5.073e-01)
(3.333, 3.453e-01)
(6.667, 2.262e-01)
(10.000, 2.067e-01)
(13.333, 1.788e-01)
(16.667, 1.764e-01)
(20.000, 1.681e-01)
(23.333, 1.689e-01)
(26.667, 1.672e-01)
(30.000, 1.675e-01)
};

\addplot 
coordinates
{
(0.000, 8.223e-01)
(3.333, 7.998e-01)
(6.667, 7.867e-01)
(10.000, 7.816e-01)
(13.333, 7.769e-01)
(16.667, 7.779e-01)
(20.000, 7.770e-01)
(23.333, 7.773e-01)
(26.667, 7.760e-01)
(30.000, 7.728e-01)
};

\legend{0 ns,10 ns,20 ns,50 ns,100 ns, 300 ns}
\end{semilogyaxis}
\end{tikzpicture}

%% file: evm_plots/doppler_sweep.tex
\begin{tikzpicture}[every node/.style={scale=0.70}]
\begin{loglogaxis}[
	scale only axis,
	separate axis lines=false, 
	xtick pos=left,
	ytick pos=left,
	ylabel={EVM$^2$},
    xlabel={Doppler Spread (Hz)},
    xmin = 50, xmax = 5000,
    ymin = 1e-5, ymax = 1e1,
    cycle list name=set1,
    label style={font=\LARGE},
    tick label style={font=\LARGE},
    legend style={font=\Large,
    			  cells={anchor=west},
    			  anchor=south east,
                  at={(0.75,0.05)}}
]

\addplot 
coordinates
{
(5.00e+01,2.08e-04)
(9.65e+01,7.44e-04)
(1.86e+02,2.77e-03)
(3.60e+02,1.10e-02)
(6.95e+02,5.05e-02)
(1.34e+03,3.52e-01)
(2.59e+03,2.24e+00)
(5.00e+03,5.33e+00)
};

\addplot
coordinates
{
(5.00e+01,2.08e-04)
(9.65e+01,7.45e-04)
(1.86e+02,2.77e-03)
(3.60e+02,1.10e-02)
(6.95e+02,5.05e-02)
(1.34e+03,3.52e-01)
(2.59e+03,2.24e+00)
(5.00e+03,5.33e+00)
};

\addplot
coordinates
{
(5.00e+01,1.35e-02)
(9.65e+01,1.29e-02)
(1.86e+02,1.40e-02)
(3.60e+02,1.38e-02)
(6.95e+02,1.48e-02)
(1.34e+03,5.55e-02)
(2.59e+03,6.12e-02)
(5.00e+03,6.13e-02)
};

\addplot 
coordinates
{
(5.00e+01,1.34e-02)
(9.65e+01,1.29e-02)
(1.86e+02,1.41e-02)
(3.60e+02,1.38e-02)
(6.95e+02,1.48e-02)
(1.34e+03,5.55e-02)
(2.59e+03,6.12e-02)
(5.00e+03,6.13e-02)
};
\legend{SC-FDE, OFDM-FDE, SC-TDE, FDM-FDCP}
\end{loglogaxis}

\begin{loglogaxis}[
    scale only axis,
    separate axis lines=false, 
    x axis line style={-},
    xmax = 1/50, xmin = 1/8000,
    ymin = 1e-18, ymax = 9e3,
    axis y line = right,
    axis x line = top,
    x axis line style=-,
    y axis line style=-,
    ytick={1,1e-2,1e-4,1e-6,1e-9,1e-12},
    ylabel = Approximate SER Floor,
    xlabel = Coherence Time (s),
    label style={font=\LARGE},
    tick label style={font=\LARGE},
    x dir=reverse
    ]

\draw[dashed] (axis cs: 2.2e-4,1e-19) -- (axis cs: 2.2e-4,1e9);
\node[align=left] (bl) at (axis cs: 5.5e-4, 1e-9) {\Large Block Duration \\ \Large $2.3 \times 10^{-4}\,$s};
\draw[->] (bl) -- (axis cs: 2.3e-4, 1e-9);
\end{loglogaxis}
\end{tikzpicture}

%% file: evm_plots/delay_sweep.tex
\begin{tikzpicture}[every node/.style={scale=0.70}]
\begin{loglogaxis}[
	scale only axis,
	separate axis lines=false, 
	xtick pos=left,
	ytick pos=left,
	ylabel={EVM$^2$},
    xlabel={Delay Spread (s)},
    ymin = 1e-5, ymax = 10,
    xmin = 1e-8, xmax = 1e-6,
    cycle list name=set1,
    label style={font=\LARGE},
    tick label style={font=\LARGE},
    legend style={font=\Large,
    			  cells={anchor=west},
    			  anchor=south east,
                  at={(0.985,0.02)}}
]

\addplot
coordinates
{
(1.000e-08, 2.226e-02)
(1.931e-08, 2.220e-02)
(3.728e-08, 2.229e-02)
(7.197e-08, 2.461e-02)
(1.389e-07, 2.810e-02)
(2.683e-07, 3.603e-02)
(5.179e-07, 5.454e-02)
(1.000e-06, 6.898e-02)
};

\addplot
coordinates
{
(1.000e-08, 2.222e-02)
(1.931e-08, 2.228e-02)
(3.728e-08, 2.226e-02)
(7.197e-08, 2.453e-02)
(1.389e-07, 2.848e-02)
(2.683e-07, 3.638e-02)
(5.179e-07, 5.465e-02)
(1.000e-06, 7.159e-02)
};

\addplot
coordinates
{
(1.000e-08, 4.019e-06)
(1.931e-08, 3.123e-04)
(3.728e-08, 4.839e-03)
(7.197e-08, 3.283e-02)
(1.389e-07, 2.512e-01)
(2.683e-07, 7.944e-01)
(5.179e-07, 1.623e+00)
(1.000e-06, 1.846e+00)
};

\addplot
coordinates
{
(1.000e-08, 4.093e-06)
(1.931e-08, 3.123e-04)
(3.728e-08, 4.819e-03)
(7.197e-08, 3.269e-02)
(1.389e-07, 2.518e-01)
(2.683e-07, 7.948e-01)
(5.179e-07, 1.625e+00)
(1.000e-06, 1.849e+00)
};

\draw[dashed] (axis cs: 2.2e-7,1e-6) -- (axis cs: 2.2e-7,8);
\node[align=left] (ts) at (axis cs: 5.25e-7, 1e-3) {\Large Symbol Period \\ \Large $2.2 \times 10^{-7}\,$s};
\draw[->] (ts) -- (axis cs: 2.3e-7, 1e-3);

\legend{ SC-FDE, OFDM-FDE, SC-TDE, FDM-FDCP}
\end{loglogaxis}

\begin{semilogyaxis}[
    scale only axis,
    separate axis lines=false, 
    xmin = 0, xmax = 2,
    ymin = 1e-18, ymax = 7e3,
    axis y line = right,
    axis x line = none,
    x axis line style=-,
    y axis line style=-,
    ytick={1,1e-2,1e-4,1e-6,1e-9,1e-12},
    ylabel = Approximate SER Floor,
    label style={font=\LARGE},
    tick label style={font=\LARGE}
    ]
\end{semilogyaxis}
\end{tikzpicture}

%% file: evm_plots/symbol_rate_sweep.tex
\begin{tikzpicture}[every node/.style={scale=0.70}]
\begin{loglogaxis}[
	scale only axis,
	separate axis lines=false, 
	xtick pos=left,
	ytick pos=left,
	ylabel={EVM},
    xlabel={Symbol Period, $T_s$ (s)},
    xmin = 2.2e-8, xmax = 2.4e-6,
    ymin = 8e-6, ymax = 3,
    cycle list name=set1,
    max space between ticks=32,
    label style={font=\LARGE},
    tick label style={font=\LARGE},
    legend style={font=\large,
    			  cells={anchor=west},
    			  anchor=south west,
                  at={(0.0125,0.0125)}}
]

\addplot 
coordinates
{
(2.222e-06, 3.334e+00)
(1.151e-06, 1.435e+00)
(5.962e-07, 3.432e-01)
(3.088e-07, 5.454e-02)
(1.599e-07, 1.373e-02)
(8.284e-08, 4.328e-03)
(4.290e-08, 1.677e-03)
(2.222e-08, 5.925e-04)
};

\addplot
coordinates
{
(2.222e-06, 3.325e+00)
(1.151e-06, 1.437e+00)
(5.962e-07, 3.428e-01)
(3.088e-07, 5.429e-02)
(1.599e-07, 1.403e-02)
(8.284e-08, 4.457e-03)
(4.290e-08, 1.485e-03)
(2.222e-08, 6.165e-04)
};

\addplot
coordinates
{
(2.222e-06, 1.559e-05)
(1.151e-06, 1.280e-03)
(5.962e-07, 1.621e-02)
(3.088e-07, 5.834e-02)
(1.599e-07, 2.551e-01)
(8.284e-08, 8.115e-01)
(4.290e-08, 1.581e+00)
(2.222e-08, 1.852e+00)
};

\addplot
coordinates
{
(2.222e-06, 1.531e-05)
(1.151e-06, 1.282e-03)
(5.962e-07, 1.621e-02)
(3.088e-07, 5.347e-02)
(1.599e-07, 2.557e-01)
(8.284e-08, 8.119e-01)
(4.290e-08, 1.584e+00)
(2.222e-08, 1.855e+00)
};

\draw[dashed] (axis cs: 1e-7,8e-6) --
(axis cs: 1e-7,3);

\node[align=left] (ds) at (axis cs: 2.5e-7, 1e-4) {\large Delay Spread \\ \large $10^{-7}\,$s};

\draw[->] (ds) -- (axis cs: 1.02e-7, 1e-4);

\legend{SC-FDE, OFDM-FDE, SC-TDE, FDM-FDCP}
\end{loglogaxis}

\begin{loglogaxis}[
    scale only axis,
    separate axis lines=false, 
    x axis line style={-},
    xmin = 1024*2e-8, xmax = 1024*2.5e-6,
    ymin = 1e-25, ymax = 2e2,
    axis y line = right,
    axis x line = top,
    x axis line style=-,
    y axis line style=-,
    ytick={1,1e-2,1e-4,1e-6,1e-9,1e-12},
    ylabel = Approximate SER Floor,
    xlabel = Block Duration (s),
    label style={font=\LARGE},
    tick label style={font=\LARGE},
    ]

\draw[dashed] (axis cs: 2e-3,1e-25) --
(axis cs: 2e-3,2e2);

\node[align=left] (ct) at (axis cs: 8.5e-4, 2e-6) {\large Coherence Time \\ \large $2 \times 10^{-3}\,$s};

\draw[->] (ct) -- (axis cs: 1.97e-3, 2e-6);

\end{loglogaxis}
\end{tikzpicture}

%% file: evm_plots/block_length_sweep.tex
\begin{tikzpicture}[every node/.style={scale=0.70}]
\begin{semilogyaxis}[
	scale only axis,
	separate axis lines=false, 
	xtick pos=left,
	ytick pos=left,
	ylabel={EVM},
    xlabel={Block Length (Samples)},
    xmin = 0, xmax = 8192,
    ymin = 1e-4, ymax = 6,
    cycle list name=set1,
    label style={font=\LARGE},
    tick label style={font=\LARGE},
    legend style={font=\large,
    			  cells={anchor=west},
    			  anchor=south east,
                  at={(0.95,0.05)}}
]

\addplot
coordinates
{
(64,2.75e-4)
(128,3.99e-4)
(256,1.27e-3)
(512,5.10e-3)
(1024,2.34e-2)
(2048,1.38e-1)
(4096,1.68e0)
(8192,2.51e0)
};

\addplot
coordinates
{
(64,2.74e-4)
(128,3.98e-4)
(256,1.27e-3)
(512,5.08e-3)
(1024,2.34e-2)
(2048,1.37e-1)
(4096,1.66e0)
(8192,2.52e0)
};

\addplot 
coordinates
{
(64,1.00e-2)
(128,1.25e-2)
(256,1.26e-2)
(512,1.42e-2)
(1024,1.40e-2)
(2048,1.59e-2)
(4096,1.88e-2)
(8192,1.94e-2)
};

\addplot
coordinates
{
(64,1.03e-2)
(128,1.26e-2)
(256,1.25e-2)
(512,1.42e-2)
(1024,1.41e-2)
(2048,1.62e-2)
(4096,1.88e-2)
(8192,1.94e-2)
};

\legend{SC-FDE, OFDM-FDE, SC-TDE, FDM-FDCP}
\end{semilogyaxis}

\begin{semilogyaxis}[
    scale only axis,
    separate axis lines=false, 
    x axis line style={-},
    xmin = 0, xmax = 1.9,
    ymin = 1e-19, ymax = 4e3,
    axis y line = right,
    axis x line = top,
    x axis line style=-,
    y axis line style=-,
    ytick={1,1e-2,1e-4,1e-6,1e-9,1e-12},
    xlabel = Block Duration (ms),
    ylabel = Approximate SER Floor,
    label style={font=\LARGE},
    tick label style={font=\LARGE},
    ]
\end{semilogyaxis}
\end{tikzpicture}